\documentclass[11pt]{article}

\usepackage{amsmath}

\newcommand{\N}{N\raise.7ex\hbox{\underline{$\circ $}}$\;$}

\begin{document}

\begin{center}

{{\bf V.V. Kisel, E.M. Ovsiyuk, V.M. Red'kov, N.G. Tokarevskaya \\ Maxwell equations  in  matrix form,  squaring
procedure, \\ separating the variables, and structure of electromagnetic solutions }\\[2mm]
Belorussian State  Pedagogical University, Minsk \\
Institute of Physics, National Academy of Sciences of Belarus\\
Belorussian Economical  University, Minsk
}

\end{center}

\begin{quotation}

The Riemann -- Silberstein -- Majorana -- Oppenheimer approach to
the Maxwell electro\-dynamics in vacuum  is investigated within the  matrix formalism.  The matrix form
of electrodynamics includes  three   real $4 \times 4$
matrices: $(-i \partial_{0} + \alpha^{j}
\partial_{j} ) \Psi (x) = 0$, where $\Psi (x) = (0 , {\bf E}(x)  +
ic {\bf B}(x) ) $. Within the  squaring procedure we construct four formal solutions of the Maxwell equations on the
base of scalar  Klein -- Fock -- Gordon solutions:
$\{ \Psi^{0} ,  \Psi^{1}, \Psi^{2} ,\Psi^{3} \} = ( i
\partial_{0} + \alpha^{1} \partial_{1} + \alpha^{2} \partial_{2} +
\alpha^{3}
\partial_{3} )\;\Phi (x)$,
 where  $\Phi (x)$ satisfies equation $\partial^{a}\partial_{a} \Phi (x) = 0$.
The problem of separating physical electromagnetic waves   in the linear space
$ \{ \lambda_{0} \Psi^{0} +  \lambda_{1}  \Psi^{1} + \lambda_{2} \Psi^{2} + \lambda_{3}\Psi^{3} \}$
is investigated, several   particular cases, plane waves and cylindrical waves,  are considered in detail.

\end{quotation}

\section{Introduction}

Special relativity arose from study of
the Maxwell equations symmetry  with respect to motion of references frames:
Lorentz  \cite{1904-Lorentz},  Poincar\'e \cite{1905-Poincare},
Einstein \cite{1905-Einstein} Naturally, an analysis of the
Maxwell equations with respect to Lorentz transformations was the
first objects of relativity theory:
 Minkowski \cite{1908-Minkowski},
 Silberstein \cite{1907-Silberstein(1), 1907-Silberstein(2)},
Marcolongo \cite{1914-Marcolongo},  Bateman  \cite{1915-Bateman},
and  Lanczos \cite{1919-Lanczos}, Gordon \cite{1923-Gordon},
Mandel'stam --  Tamm
\cite{1925-Mandel'stam, 1925-Tamm(1), 1925-Tamm(2)}.

After Dirac  \cite{1928-Dirac} discovery of the relativistic equation for a particle with
spin 1/2  much work was done to study
spinor and vectors within the Lorentz group theory: M\"{o}glich
\cite{1928-Moglich},  Ivanenko -- Landau
\cite{1928-Ivanenko-Landau}, Neumann \cite{1929-Neumann}, van der
Waerden \cite{1929-Waerden}, Juvet \cite{1930-Juvet}. As was shown
any quantity which transforms linearly under Lorentz
transformations is a spinor. For that reason spinor quantities are
considered as fundamental in quantum field theory and basic
equations for such quantities should be written in a spinor form.
A spinor formulation of Maxwell equations was studied by Laporte
and Uhlenbeck \cite{1931-Laporte}, also see Rumer
\cite{1936-Rumer}. In 1931,   Majorana \cite{1931-Majorana} and
Oppenheimer \cite{1931-Oppenheimer} proposed to consider the
Maxwell theory of electromagnetism as the wave mechanics of the
photon. They introduced a  complex 3-vector wave function
satisfying the massless Dirac-like equations. Before Majorana  and
Oppenheimer,    the most crucial steps were made by Silberstein
\cite{1907-Silberstein(1)}, he showed the possibility to have
formulated Maxwell equation in term of complex 3-vector entities.
Silberstein in his second paper  \cite{1907-Silberstein(2)} writes
that the complex form of Maxwell equations has been known before;
he refers there to the second volume of the lecture notes on the
differential equations of mathematical physics by B. Riemann that
were edited and published by H. Weber in 1901 \cite{1901-Weber}.
 This not widely used fact  is  noted by Bialynicki-Birula   \cite{1994-Bialynicki-Birula}).

Maxwell equations in the  matrix Dirac-like  form considered
during long time by many authors:
Luis de Broglie
\cite{1934-Broglie(1), 1934-Broglie(2), 1939-Broglie, 1940-Broglie},
 Petiau \cite{1936-Petiau}, Proca \cite{1936-Proca, 1946-Proca},
Duffin \cite{1938-Duffin}, Kemmer
\cite{1939-Kemmer, 1943-Kemmer, 1960-Kemmer},  Bhabha
\cite{1939-Bhabha}, Belinfante
\cite{1939-Belinfante(1), 1939-Belinfante(2)}, Taub
\cite{1939-Taub},  Sakata  -- Taketani \cite{1940-Sakata},
Schr\"{o}dinger
\cite{1943-Schrodinger(1), 1943-Schrodinger(2)}, Heitler
\cite{1943-Heitler},
\cite{1946-Harish-Chandra(1), 1946-Harish-Chandra(2)},
Mercier \cite{1949-Mercier}, Imaeda \cite{1950-Imaeda},   Fujiwara
\cite{1953-Fujiwara}, Ohmura  \cite{1956-Ohmura},
  Borgardt \cite{1956-Borgardt, 1958-Borgardt},  Fedorov \cite{1957-Fedorov},  Kuohsien \cite{1957-Kuohsien},
Bludman   \cite{1957-Bludman}, Good \cite{1957-Good},  Moses
\cite{1958-Moses, 1959-Moses, 1973-Moses}, Lomont
\cite{1958-Lomont}, Bogush -- Fedorov \cite{1962-Bogush-Fedorov},
Sachs -- Schwebel  \cite{1962-Sachs-Schwebel},   Ellis
\cite{1964-Ellis}, Oliver \cite{1968-Oliver}, Beckers -- Pirotte
\cite{1968-Beckers},  Casanova \cite{1969-Casanova},  Carmeli
\cite{1969-Carmeli}, Bogush \cite{1971-Bogush}, Lord
\cite{1972-Lord}, Weingarten \cite{1973-Weingarten},   Mignani --
Recami --  Baldo  \cite{1974-Recami1}, Newman \cite{1973-Newman},
\cite{1974-Frankel},  \cite{1975-Jackson}, Edmonds
\cite{1975-Edmonds}, Silveira \cite{1980-Silveira};
the interest to the
Majorana -- Oppenheimer formulation of electrodynamics has grown in
recent years:
 Jena --
Naik --  Pradhan \cite{1980-Jena},  Venuri  \cite{1981-Venuri},
Chow \cite{1981-Chow}, Fushchich -- Nikitin
\cite{1983-Fushchich}, Cook
\cite{1982-Cook(1), 1982-Cook(2)}, Giannetto
\cite{1985-Giannetto},   Y\'epez, Brito -- Vargas
\cite{1988-Nunez}, Kidd --  Ardini --  Anton  \cite{1989-Kidd},
Recami \cite{1990-Recami}, Krivsky --  Simulik
\cite{1992-Krivsky}, Inagaki \cite{1994-Inagaki},
Bialynicki-Birula
\cite{1994-Bialynicki-Birula, 1996-Bialynicki-Birula, 2005-Birula},
Sipe \cite{1995-Sipe}, \cite{1996-Ghose}, Esposito
\cite{1998-Esposito}, Dvoeglazov \cite{1998-Dvoeglazov} (see a big
list of  relevant references therein)-\cite{2001-Dvoeglazov},
Gersten  \cite{1998-Gersten}, Kanatchikov \cite{2000-Kanatchikov},
Gsponer \cite{2002-Gsponer}, Ivezic
\cite{2001-Ivezic(1),2002-Ivezic(2),    2002-Ivezic(3), 2002-Ivezic,  2003-Ivezic, 2005-Ivezic(1),
2005-Ivezic(2), 2005-Ivezic(3), 2006-Ivezic},
Donev --  Tashkova.
\cite{2004-Donev(1), 2004-Donev(2), 2004-Donev(3)}.

Our treatment will be with a quite definite  accent:
 the main attention is given to possibilities given by the matrix approach
 for explicit constructing electromagnetic solutions of the Maxwell equations.
 In vacuum case, the matrix form includes  three   real $4 \times 4$
matrices $\alpha^{j}$:
\begin{eqnarray}
(-i \partial_{0} + \alpha^{j} \partial_{j} ) \Psi (x) = 0 \; ,
\nonumber
\end{eqnarray}

\noindent
where $\Psi (x) = (0 , {\bf E}(x)  + ic {\bf B}(x) ) $.
With the use of squaring procedure one  may construct four formal solutions of the Maxwell equations on the base of
scalar solution of  the Klein -- Fock -- Gordon  equation:
\begin{eqnarray}
\{ \Psi^{0} ,  \Psi^{1}, \Psi^{2} ,\Psi^{3} \} = ( i  \partial_{0}
+ \alpha^{1} \partial_{1} + \alpha^{2} \partial_{2} + \alpha^{3}
\partial_{3} )\;\Phi (x)\;, \qquad \partial^{a}\partial_{a} \Phi (x) = 0 \; .
\nonumber
\end{eqnarray}

\noindent
The problem of separating physical electromagnetic solutions  in the linear space
$ \{ \lambda_{0} \Psi^{0} +  \lambda_{1}  \Psi^{1} + \lambda_{2} \Psi^{2} + \lambda_{3}\Psi^{3} \}
 $
  is investigated. Several   particular cases  are considered in detail.

\section{  Complex matrix form of Maxwell theory in vacuum }

Let us start with Maxwell equations in  vacuum
\cite{1941-Stratton, 1961-Panofsky-Phillips, 1975-Jackson, 1973-Landau}
(with the use of usual notation for  current 4-vector $
j^{a} = (\rho, {\bf J} /c) \; ,  \; c^{2} = 1 /
\epsilon_{0}\mu_{0}  $ ):
\begin{eqnarray}
 \mbox{div} \; c{\bf B} = 0 \; , \qquad \mbox{rot}
\;{\bf E} = -{\partial c {\bf B} \over \partial ct} \; , \nonumber
\\
 \mbox{div}\; {\bf E} = {\rho \over  \epsilon_{0}} , \qquad
 \mbox{rot} \; c{\bf B} =  {{\bf j} \over \epsilon_{0}}  +
   {\partial {\bf E} \over \partial ct} \; ,
\label{1.2a}
\end{eqnarray}

Let us introduce 3-dimensional complex vector
$
\psi^{k} =   E^{k} + i c B^{k} \; ,
$ with the help of which the above equations can be
combined into (see  Silberschtein
\cite{1907-Silberstein(1), 1907-Silberstein(2)},
 Bateman  \cite{1915-Bateman}, Majorana \cite{1931-Majorana},  Oppenheimer \cite{1931-Oppenheimer},
 and many others)
\begin{eqnarray}
\partial_{1}\Psi ^{1} + \partial_{2}\Psi ^{0} + \partial_{3}\Psi ^{3} =
j^{0} / \epsilon_{0}  \; ,\;\; -i\partial_{0} \psi^{1} +
(\partial_{2}\psi^{3} -
\partial_{3}\psi^{2}) = i\; j^{1} / \epsilon_{0} \; ,
\nonumber
\\
-i\partial_{0} \psi^{2} + (\partial_{3}\psi^{1} -
\partial_{1}\psi^{3}) = i\; j^{2} / \epsilon_{0} \; , \;
-i\partial_{0} \psi^{3} + (\partial_{1}\psi^{2} -
\partial_{2}\psi^{1}) =  i\; j^{3} / \epsilon_{0} \; .
\label{1.3b}
\end{eqnarray}

\noindent  let $x_{0}=ct, \; \partial_{0} = c \partial_{t}$.
These four relations can be rewritten in a  matrix form using a
4-dimensional column  $\psi$ with one additional zero-element
\cite{1983-Fushchich, Book}:
\begin{eqnarray}
(-i \partial_{0} + \alpha^{j} \partial_{j} ) \Psi =J \; , \qquad
\Psi = \left | \begin{array}{c} 0 \\\psi^{1} \\\psi^{2} \\
\psi^{3}
\end{array} \right | \; , \qquad J=
{1 \over \epsilon_{0}} \; \left | \begin{array}{c} j^{0} \\ i\;
j^{1} \\ i\; j^{2} \\ i \; j^{3}
\end{array} \right | \; ,
\nonumber
\\
\alpha^{1} = \left | \begin{array}{rrrr}
0 & 1  &  0  & 0  \\
-1 & 0  &  0  & 0  \\
0 & 0  &  0  & -1  \\
0 & 0  &  1  & 0
\end{array}  \right |, \qquad
\alpha^{2} = \left | \begin{array}{rrrr}
0 & 0  &  1  & 0  \\
0 & 0  &  0  & 1  \\
-1 & 0  &  0  & 0  \\
0 & -1  & 0  & 0
\end{array}  \right |, \qquad
\alpha^{3} = \left | \begin{array}{rrrr}
0 & 0  &  0  & 1  \\
0 & 0  & -1  & 0  \\
0 & 1  &  0  & 0  \\
-1 & 0  &  0  & 0
\end{array}  \right |
\nonumber
\\
(\alpha^{1})^{2} = -I , \qquad  (\alpha^{2})^{2} = -I ,  \qquad
(\alpha^{2})^{2} = -I , \nonumber
\\
\alpha^{1} \alpha^{2}= - \alpha^{2} \alpha^{1} =  \alpha^{3} \;,
\qquad \alpha^{2} \alpha^{3} = - \alpha^{3} \alpha^{2} =
\alpha^{1}\;, \qquad \alpha^{3} \alpha^{1} = - \alpha^{1}
\alpha^{3} = \alpha^{2}\;. \label{1.10}
\end{eqnarray}

\section{ Method to  construct   electromagnetic   solutions from
 scalar ones}

The above  matrix form of Maxwell theory :
\begin{eqnarray}
 (-i \partial_{0} + \alpha^{j} \partial_{j} ) \Psi =0 \; , \qquad
\Psi = \left | \begin{array}{c} 0 \\\psi^{1} \\\psi^{2} \\
\psi^{3}
\end{array} \right | \; .
 \label{3.1}
\end{eqnarray}

\noindent permits us to develop a simple method of finding
solutions of Maxwell equations on the  base of known solutions of
the scalar massless  equation by  Klein -- Fock -- Gordon. Indeed, in
virtue of the  above commutative  relations  we have  an operator
identity
\begin{eqnarray}
 (-i \partial_{0} + \alpha^{1} \partial_{1} + \alpha^{2}
\partial_{2} + \alpha^{3} \partial_{3} )\; (-i \partial_{0} -
\alpha^{1} \partial_{1} - \alpha^{2} \partial_{2} - \alpha^{3}
\partial_{3} )\;=
 ( - \partial^{2}_{0} + \partial^{2} _{1} +  \partial^{2}_{2} +
\partial^{3}_{3} )\; .
\nonumber
\end{eqnarray}

\noindent Therefore,  taking any special   scalar
solution
$( - \partial^{2}_{0} + \partial^{2} _{1} +  \partial^{2}_{2} +
\partial^{2}_{3} )\; \Phi (x) = 0 \; ,
$
 one can immediately  construct four  solutions of
the  Maxwell  equation:
\begin{eqnarray}
(-  i \partial_{0} + \alpha^{1} \partial_{1} + \alpha^{2}
\partial_{2} + \alpha^{3} \partial_{3} )\;\Psi^{a} =0 \; ,
\nonumber
\label{3.3}
\end{eqnarray}

\noindent where  $\Psi^{a}$ are columns of the  matrix
\begin{eqnarray}
( i  \partial_{0}
+ \alpha^{1} \partial_{1} + \alpha^{2} \partial_{2} + \alpha^{3}
\partial_{3} )\;\Phi (x) = \qquad \qquad
\nonumber
\\
=
\left | \begin{array}{rrrr}
i\partial_{0}\Phi &  \quad  \partial_{1}\Phi  &  \quad  \partial_{2}\Phi  &  \quad  \partial_{3} \Phi  \\
-\partial_{1}\Phi &  \quad i\partial_{0}\Phi  &  \quad -\partial_{3}\Phi  &  \quad  \partial_{2} \Phi \\
-\partial_{2}\Phi &  \quad  \partial_{3}\Phi  &  \quad i\partial_{0}\Phi  &  \quad -\partial_{1}\Phi  \\
-\partial_{3}\Phi &  \quad -\partial_{2}\Phi  &  \quad  \partial_{1}\Phi  &  \quad i\partial_{0}\Phi
\end{array}  \right | =\{ \Psi^{0} ,  \Psi^{1}, \Psi^{2} ,\Psi^{3} \}
\; .
\label{3.4}
\end{eqnarray}

Thus,  we have four formal solutions of the free Maxwell equations (let $F_{a} (x) = \partial_{a} \Phi (x)$):
\begin{eqnarray}
\{ \Psi^{0} ,  \Psi^{1}, \Psi^{2} ,\Psi^{3} \} =
\left | \begin{array}{rrrr}
i F_{0}  &  \quad  F_{1} &  \quad  F_{2}  &  \quad  F_{3}   \\
-F_{1}   &  \quad iF_{0} &  \quad -F_{3}  &  \quad  F_{2}   \\
-F_{2}   &  \quad  F_{3} &  \quad iF_{0}  &  \quad -F_{1}   \\
-F_{3}   &  \quad -F_{2} &  \quad  F_{1}  &  \quad iF_{0}
\end{array}  \right | \; .
\label{3.5}
\end{eqnarray}

\section{ Electromagnetic plane waves from scalar ones }

Let us specify this method for  an elementary   example,  starting from
a scalar plane wave propagating along the axis $z$:
\begin{eqnarray}
\Phi =  A  \sin (\omega t - k z ) = A  \sin (k_{0} x_{0} - k_{3}
x_{3} ) = A \cos \varphi  \; .
\label{3.6}
\end{eqnarray}

\noindent The  recipe (\ref{3.4})  gives
\begin{eqnarray}
\{ \Psi^{a} \}  =  A \left | \begin{array}{rrrr}
i k_{0} & 0  &  0  & - k_{3}  \\
0 & i k_{0}  &  k_{3}  & 0  \\
0 & - k_{3}  &  i k_{0}  & 0  \\
 k_{3} & 0  &  0  & i k_{0}
\end{array}  \right |  \cos \varphi
\; .
\label{3.7}
\end{eqnarray}

\noindent Because the left and the right  columns have
non-vanishing zero-component, they cannot  represent any  real
solutions  of the  Maxwell  equations. However,   two
remaining  seem to be  suitable ones (the factor $ A \cos \varphi  $
is omitted):
\begin{eqnarray}
\Psi ^{I} =   \left | \begin{array}{r} 0 \\ ik_{0} \\ - k_{3}
\\0
\end{array}  \right |  =
\left | \begin{array}{c}
0 \\ E^{1} + i cB^{1} \\ E^{2} + i cB^{2} \\
E^{3} + ci B^{3}  \end{array} \right | \; , \qquad
\Psi ^{II} =  A  \left | \begin{array}{r} 0 \\ k_{3}  \\  ik_{0}
\\0
\end{array}  \right |  =
\left | \begin{array}{c}
0 \\ E^{1} + i cB^{1} \\ E^{2} + i cB^{2} \\
E^{3} + ci B^{3}  \end{array} \right |  \; .
\nonumber
\end{eqnarray}

\noindent Thus, we have two  wave solutions:
\begin{eqnarray}
I  \qquad      E^{2} = - k_{3}  A         \;   \cos \varphi  \; ,
      \qquad     B^{1} = {k_{0} \over c} A \;   \cos \varphi  \; ,
\nonumber
\\[2mm]
II \qquad E^{1} = k_{3}A\;  \cos  \varphi \;
, \qquad B^{2} ={ k_{0} \over c} A\;  \cos \varphi  \; .
\label{3.9}
\end{eqnarray}

\noindent In both cases, the ratio $E / B$  equals to  the speed
of  light:
\begin{eqnarray}
{E \over B } = {k_{3} \over k_{0} /c } = c \; .
\nonumber
\end{eqnarray}

\noindent Besides, both waves have are   clockwise polarized:
\begin{eqnarray}
I \qquad  ({\bf E} \times {\bf B} )= +
 {\bf e}_{3}\;  {k_{3}k_{0} \over c} A^{2} \cos^{2} \varphi \; ,
\nonumber
\\
II \qquad  ({\bf E} \times {\bf B} )= +
 {\bf e}_{3}\;  {k_{3}k_{0} \over c} A^{2}  \cos^{2} \varphi \; .
\label{3.10b}
\end{eqnarray}

\noindent Two waves are linearly independent  and
orthogonal to each other:
\begin{eqnarray}
{\bf E}^{I} {\bf E}^{II}  =0 \; , \qquad  {\bf B}^{I} {\bf
B}^{II}  =0 \; .
\nonumber
\end{eqnarray}

In the same manner we can solve  a more general problem of
constructing plane wave solutions with arbitrary wave  vector
${\bf k}$. Let  us start with a scalar wave
\begin{eqnarray}
\Phi = A \;
 \sin (k_{0} x_{0} - k_{i} x_{i}) = A \; \cos \varphi  \; .
\label{3.11a}
\end{eqnarray}

\noindent
The matrix of solutions (\ref{3.4})  will take the form
\begin{eqnarray}
\{ \Psi^{0} ,  \Psi^{1}, \Psi^{2} ,\Psi^{3} \} = A \left | \begin{array}{rrrr}
ik_{0} & -k_{1}   &  -k_{2}  & -k_{3}  \\
k_{1} & ik_{0}  &  +k_{3}  & -k_{2}  \\
k_{2} & -k_{3}  &  ik_{0}  & +k_{1}  \\
k_{3} & +k_{2}  &  -k_{1}  & ik_{0}
\end{array}  \right | \cos \varphi  \; .
\label{3.11b}
\end{eqnarray}

\noindent
We have four formal solutions $\Psi^{a}$, but  they  cannot
be  regarded as  physical  because each of them has  a
non-vanishing  zero-component. However, we can use linearity of
the Maxwell  equation and combine elementary columns in (\ref{3.11b})
 with any coefficients. In this way we  are  able to construct  physical solutions.
For shortness  let us  omit the  factor  $A\;\sin \varphi $  and  operate only with the columns of the matrix
\begin{eqnarray}
\{ \Psi^{0} ,  \Psi^{1}, \Psi^{2} ,\Psi^{3} \} = \left |
\begin{array}{rrrr}
ik_{0} & -k_{1}   &  -k_{2}  & -k_{3}  \\
k_{1} & ik_{0}  &  +k_{3}  & -k_{2}  \\
k_{2} & -k_{3}  &  ik_{0}  & +k_{1}  \\
k_{3} & +k_{2}  &  -k_{1}  & ik_{0}
\end{array}  \right | \; .
\nonumber
\end{eqnarray}

\noindent Taking into account  properties of  the wave  vector
\begin{eqnarray}
k_{0}^{2} - {\bf k}^{2} = 0 \; \qquad \Longrightarrow \qquad {\bf
k} = k_{0} {\bf n} \; , \; {\bf n}^{2} = 1 \; ,
\label{3.13}
\end{eqnarray}

\noindent previous  matrix can be rewritten as follows (the common
factor  $k_{0}$  is  omitted)
\begin{eqnarray}
\{ \Psi^{0} ,  \Psi^{1}, \Psi^{2} ,\Psi^{3} \} = k_{0} \left |
\begin{array}{cccc}
i & -n_{1}   &  -n_{2}  & -n_{3}  \\
n_{1} & i  &  +n_{3}  & -n_{2}  \\
n_{2} & -n_{3}  &  i  & +n_{1}  \\
n_{3} & +n_{2}  &  -n_{1}  & i
\end{array}  \right | \; \sim
\left | \begin{array}{cccc}
i & -n_{1}   &  -n_{2}  & -n_{3}  \\
n_{1} & i  &  +n_{3}  & -n_{2}  \\
n_{2} & -n_{3}  &  i  & +n_{1}  \\
n_{3} & +n_{2}  &  -n_{1}  & i
\end{array}  \right | \;.
\nonumber
\end{eqnarray}

\noindent  First,
with the  help of the column (0) let us produce  a zero at the
first component of the columns
  (1) -- (2) -- (3):
\begin{eqnarray}
\{ \Psi^{0} ,  \Psi^{1}, \Psi^{2} ,\Psi^{3} \}  \sim \left |
\begin{array}{llll}
i      &   0                      &  0                    & 0   \\
n_{1}  &  -in_{1}n_{1} + i        &  -in_{2}n_{1} +n_{3}  &  -in_{3}n_{1} -n_{2}  \\
n_{2}  &   -in_{1}n_{2}  -n_{3}   &  -in_{2}n_{2} +i      &  -in_{3}n_{2} +n_{1}  \\
n_{3}  &   -in_{1} n_{3} +n_{2}   &  -in_{2}n_{3} -n_{1}  &
-in_{3}n_{3}  +i
\end{array}  \right | \;.
\nonumber
\end{eqnarray}

\noindent Noting that the column (3)  is a linear combination of
the columns (1) and (2):
\begin{eqnarray}
 i\; n_{2} \;(1) - i\;  n_{1} \;(2)  = (3) \; ;
\nonumber
\end{eqnarray}

\noindent in other words,  solution (3) is a linear combination
of  (1) and  (2). Therefore from the above  it follows
(multiplying the columns (1) and (2) by
imaginary $i$):
\begin{eqnarray}
\{ \Psi^{0} ,  \Psi^{1}, \Psi^{2} ,\Psi^{3} \}  \sim \left |
\begin{array}{llll}
i      &   0                    &  0                    & 0  \\
n_{1}  &  n_{1}^{2}-1           &  n_{2}n_{1} + in_{3}  & 0  \\
n_{2}  &  n_{1}n_{2} -i n_{3}   &  n_{2}n_{2} -1        & 0  \\
n_{3}  &  n_{1} n_{3} +in_{2}   &  n_{2}n_{3} -in_{1}   & 0
\end{array}  \right | \;.
\label{3.14}
\end{eqnarray}

\noindent Thus, two physical Maxwell solutions  are
\begin{eqnarray}
\Psi^{I} = \left | \begin{array}{c}
0  \\
n_{1}^{2}-1   \\
n_{1}n_{2} -i n_{3}  \\
n_{1} n_{3} +in_{2}
\end{array} \right |  =
\left | \begin{array}{c}
0 \\ E^{1} + i cB^{1} \\ E^{2} + i cB^{2} \\
E^{3} + ci B^{3}  \end{array} \right |
 \; , \qquad
\Psi^{II} = \left | \begin{array}{c}
0  \\
n_{2}n_{1} + in_{3}  \\
n_{2}n_{2} -1 \\
n_{2}n_{3} -in_{1}
\end{array} \right |  =
\left | \begin{array}{c}
0 \\ E^{1} + i cB^{1} \\ E^{2} + i cB^{2} \\
E^{3} + ci B^{3}  \end{array} \right |
 \; ,
\label{3.15}
\end{eqnarray}

\noindent or differently (the factor $A\; \cos \varphi $ is omitted)
\begin{eqnarray}
I \qquad \;\; {\bf E} =   ( n_{1}^{2}-1 ,\; n_{1}n_{2}      , \;n_{1}
n_{3}   ) \;    \; , \qquad
 c {\bf B} = ( 0  ,  \;- n_{3}   ,\; n_{2}  )   \;  ;
\nonumber
\\
II \qquad {\bf E} = ( n_{2}n_{1} ,\; n_{2}n_{2} -1     ,
\;n_{2}n_{3}  )  \; , \qquad
 c {\bf B} = (n_{3}   , \;   0 ,\; -n_{1}   )   \;  .
\label{3.16}
\end{eqnarray}

\noindent It is the matter of simple calculation  to verify  the
identity for amplitudes:
 $cB=E$. Also,  both these waves  are  clockwise polarized:
\begin{eqnarray}
{\bf E}^{I} \times  {\bf B}^{I}  = \left |
\begin{array}{rrr}
{\bf e}_{1}  &  {\bf e}_{2}  & {\bf e}_{3} \\
n_{1}^{2}-1  & n_{1}n_{2}    & n_{1} n_{3} \\
 0           & - n_{3}       &  n_{2}
\end{array} \right | =
(1 - n_{1}^{2} ) \; (n_{1} {\bf e}_{1}  + n_{2}  {\bf e}_{2}  +
n_{3}  {\bf e}_{1} )  \; \sim {\bf k} \; ,
\nonumber
\\
{\bf E}^{II} \times  {\bf B}^{II}  = \left |
\begin{array}{rrr}
{\bf e}_{1}  &  {\bf e}_{2}  & {\bf e}_{3} \\
n_{2}n_{1}   & n_{2}n_{2} -1 & n_{2}n_{3}  \\
n_{3}        &  0            &  -n_{1}
\end{array} \right | =
(1 - n_{2}^{2} ) \; (n_{1} {\bf e}_{1}  + n_{2}  {\bf e}_{2}  +
n_{3}  {\bf e}_{1} )  \; \sim {\bf k} \; .
\label{3.17}
\end{eqnarray}

\noindent Besides, they are independent solutions (but not
orthogonal ones)
\begin{eqnarray}
{\bf E}^{I}   {\bf E}^{II} = - n_{1} n_{2} \; , \qquad {\bf
B}^{I}   {\bf B}^{II} = - n_{1} n_{2} \; .
\label{3.18}
\end{eqnarray}

\section{Dual symmetry of Maxwell equations
}

Let us consider the known dual symmetry in
matrix formalism:
\begin{eqnarray}
(-i\partial_{0} + \alpha^{j} \partial_{j} )\; \left |
\begin{array}{c} 0 \\ {\bf E} + i c{\bf B}
\end{array} \right | = 0
\; .
\nonumber
\end{eqnarray}

\noindent It is evident that there exists a simple transform,
multiplication by imaginary $i$, with the following properties:
\begin{eqnarray}
\Psi^{D} = i \; \Psi  \; , \qquad  \Psi = -i\;  \Psi^{D} \;,
\qquad (-i\partial_{0} + \alpha^{j} \partial_{j} )\; \Psi^{D} = 0
\; , \qquad
\nonumber
\\
 \left | \begin{array}{c}
0 \\ {\bf E}^{D} + i c{\bf B}^{D}
\end{array} \right |=
\left | \begin{array}{c} 0 \\  i{\bf E} - c{\bf B}
\end{array} \right |  \; , \qquad
 {\bf E}^{D} = - c{\bf B}  \; , \qquad  c{\bf B}^{D} =  +{\bf E}\; .
\label{4.1b}
\end{eqnarray}

It is the dual transformation of the electromagnetic field. Several points should be clarified.
First, this transformation is not  a symmetry operation in presence of
external sources. In this case we have
\begin{eqnarray}
(-i\partial_{0} + \alpha^{j} \partial_{j} )\; \left |
\begin{array}{c} 0 \\ {\bf E} + i c{\bf B}
\end{array} \right | = {1 \over \epsilon_{0}} \; \left | \begin{array}{c}
\rho  \\ i{\bf j}
\end{array} \right |
\label{4.2a}
\end{eqnarray}

\noindent and further
\begin{eqnarray}
\Psi^{D} = i \; \Psi  \; , \qquad  \Psi = -i\;  \Psi^{D} \;,
\qquad (-i\partial_{0} + \alpha^{j} \partial_{j} )\; \Psi^{D} = {1
\over \epsilon_{0}} \; \left | \begin{array}{c} i \rho  \\ -{\bf
j}
\end{array} \right |
 ,
\nonumber
\\
 \left | \begin{array}{c}
0 \\ {\bf E}^{D} + i c{\bf B}^{D}
\end{array} \right |= \left | \begin{array}{c} 0 \\  i{\bf E} - c{\bf B}
\end{array} \right |  \; ,
\qquad
  {\bf E}^{D} = - c{\bf B}  \; , \qquad  c{\bf B}^{D} =  +{\bf E}\; .
\label{4.2b}
\end{eqnarray}

\noindent
Second, to save the situation one can   extend the Maxwell equations
by introducing  magnetic sources:
\begin{eqnarray}
(-i\partial_{0} + \alpha^{j} \partial_{j} )\;
 \left |
\begin{array}{c} 0 \\ {\bf E} + i c{\bf B}
\end{array} \right | = {1 \over \epsilon_{0}} \;
\left | \begin{array}{c} \rho_{e}  + i \rho_{m} \\ i{\bf j}_{e} +
{\bf j}_{m}
\end{array} \right |\; ,
\label{4.3}
\end{eqnarray}

\noindent  which permits us to consider the dual transformation as
a symmetry:
\begin{eqnarray}
\Psi^{D} = i \; \Psi  \; , \qquad  \Psi = -i\;  \Psi^{D} \;,
\qquad
(-i\partial_{0} + \alpha^{j} \partial_{j} )\; \Psi^{D} = {1 \over
\epsilon_{0}} \; \left | \begin{array}{c}
 - \rho_{m} +i \rho_{e}    \\  i {\bf j}_{m} -{\bf j}_{e}
\end{array} \right |
 ,
\nonumber
\\
 \left | \begin{array}{c}
0 \\ {\bf E}^{D} + i c{\bf B}^{D}
\end{array} \right |=  \left | \begin{array}{c} 0 \\  i{\bf E} - c{\bf B}
\end{array} \right |  \; ,
\qquad
  {\bf E}^{D} = - c{\bf B}  \; , \qquad  c{\bf B}^{D} =  +{\bf E}\; ,
\nonumber
\\
\rho^{D}_{e} = -\rho_{m} , \qquad {\bf j}^{D}_{e} = + {\bf j}_{m}
\; ,
\qquad
\rho ^{D}_{m} = + \rho_{e} \; , \qquad  {\bf j}^{D}_{m} = - {\bf
j}_{e} \; .
\label{4.4}
\end{eqnarray}

\noindent
In real form  eqs.  (\ref{4.4}) will look

\begin{eqnarray}
\partial_{0}  {\Psi - \Psi^{*} \over 2i } + \alpha^{j} \partial_{j} { \Psi +
   \Psi^{*}\over 2}  = {J +J^{*} \over 2} \; ,
\nonumber
\\[2mm]
 \partial_{0} { \Psi + \Psi^{*} \over 2}   + \alpha^{j} \partial_{j} { \Psi -
     \Psi^{*} \over 2i}  = {J -J^{*} \over 2i} \; ;
\nonumber
\\
Re (J) = {1 \over  \epsilon_{0}} \left | \begin{array}{c} \rho_{e}
\\  {\bf j}_{m}
\end{array} \right |\; , \qquad
Im (J) = {1 \over  \epsilon_{0}} \left | \begin{array}{c}
 \rho_{m} \\ {\bf j}_{e}
\end{array} \right |\; ,
\nonumber
\end{eqnarray}

\noindent that is
\begin{eqnarray}
\partial_{0}  B + \alpha^{j} \partial_{j} E = \mbox{Re}\; (J) \; , \qquad
\partial_{0}  E + \alpha^{j} \partial_{j} B = \mbox{Im}\;  (J) \; .
\label{4.5}
\end{eqnarray}

\noindent Eqs.  (\ref{4.5}) in  vector notation read
\begin{eqnarray}
 \mbox{div}\; {\bf E} = {\rho _{e} \over  \epsilon_{0}} , \qquad
 \mbox{rot} \;{\bf E} = + {{\bf j}_{m} \over \epsilon_{0}} -{\partial c {\bf B} \over \partial ct} \; ,
\nonumber
\\
 \mbox{div} \; c{\bf B} = {\rho _{m} \over  \epsilon_{0}} \; , \qquad
 \mbox{rot} \; c{\bf B} =  {{\bf j}_{e} \over \epsilon_{0}}  +
   {\partial {\bf E} \over \partial ct} \; .
\label{4.6}
\end{eqnarray}

Let us turn again to the Maxwell equation without sources and
examine the action of this transformation on a plane
electromagnetic wave along the axis  $z$ (see (\ref{3.9})):
\begin{eqnarray}
I \qquad E^{2} = - k_{3}  A         \;   \cos (k_{0} x_{0} -
k_{3} x_{3} ) \; , \qquad
     B^{1} = {k_{0} \over c} A \;   \cos (k_{0} x_{0} - k_{3} x_{3} ) \; .
\nonumber
\end{eqnarray}

\noindent After the dual transformation it becomes
 \begin{eqnarray}
 E^{1}_{D} = - k_{0}  A \;   \cos (k_{0} x_{0} - k_{3} x_{3} )\; ,
\qquad
  cB^{2}_{D} = - k_{3}  A         \;   \cos (k_{0} x_{0} - k_{3} x_{3} ) \; ,
\nonumber
\end{eqnarray}

\noindent It should be noted that the dual solution coincides with
the wave of the type  II according to  (\ref{3.9}):
\begin{eqnarray}
II \qquad E^{1} = k_{3}A\;  \cos (k_{0} x_{0} - k_{3} x_{3} ) \;
, \qquad
\qquad B^{2} ={ k_{0} \over c}A\;  \cos (k_{0} x_{0} - k_{3} x_{3}
) \; .
\nonumber
\end{eqnarray}

\noindent   In other words, the  dual symmetry provides us with
the  possibility to construct  a new linearly independent solution
on the base of the  known  one.

One addition should be made: the well-known continuous dual
symmetry looks as a phase transfor\-mation over complex variables:
\begin{eqnarray}
e^{i \chi}\; ({\bf E} + ic\; {\bf B}) \; ,\qquad e^{-i \chi}\; (i \; {\bf
j}_{e} +  {\bf j}_{m}) \; , \qquad e^{-i \chi}\; (\rho_{e} + i \; \rho_{m})
\;.
\label{4.7}
\end{eqnarray}

\section{  On  separating physical solutions of the  Maxwell  equations \\ (real-valued scalar function $\Phi $ ) }

Let us consider four  formal solutions of the Maxwell equations
\begin{eqnarray}
\{ \Psi^{0} ,  \Psi^{1}, \Psi^{2} ,\Psi^{3} \} =
\left | \begin{array}{rrrr}
i F_{0}  &  \quad  F_{1} &  \quad  F_{2}  &  \quad  F_{3}   \\
 -F_{1}  &  \quad iF_{0} &  \quad -F_{3}  &  \quad  F_{2}  \\
 -F_{2}  &  \quad  F_{3} &  \quad iF_{0}  &  \quad -F_{1}   \\
 -F_{3}  &  \quad -F_{2}  &  \quad F_{1}  &  \quad iF_{0}
\end{array}  \right | \; , \qquad  F_{a} (x) = \partial_{a} \Phi (x) \; .
\label{5.1}
\end{eqnarray}

\noindent Physical solutions should be
associated with the  following structure:
\begin{eqnarray}
 \left |
\begin{array}{c} 0 \\ {\bf E} + i c{\bf B}
\end{array} \right |
\nonumber
\end{eqnarray}

\noindent when zero-component  of the $ 4 \times 1$ column vanishes.

Let the function  $\Phi (x)$ be  taken as  real-valued.
We should find all  possible solutions to the following equation:
\begin{eqnarray}
\lambda _{0} \Psi^{0}  +  \lambda _{1} \Psi^{1} + \lambda _{2} \Psi^{2} + \lambda _{3}\Psi^{3}  =
\left |
\begin{array}{c} 0 \\ {\bf E} + i c{\bf B}
\end{array} \right |.
\label{5.2}
\end{eqnarray}

\noindent
Let us separate real and imaginary parts in $\lambda_{c}: \;  \lambda_{c} = a_{c} + ib_{c} $. Relation
(\ref{5.2}) takes the form
\begin{eqnarray}
(a_{0} + ib_{0}) iF_{0} + (a_{1} + ib_{1}) F_{1} + (a_{2} + ib_{2}) F_{2} + (a_{3} + ib_{3}) F_{3} = 0 \; , \qquad \qquad
\nonumber
\\
-(a_{0} + ib_{0}) F_{1} + (a_{1} + ib_{1}) iF_{0} - (a_{2} + ib_{2}) F_{3} + (a_{3} + ib_{3}) F_{2} = E_{1} + i cB_{1} \; ,
\nonumber
\\
-(a_{0} + ib_{0}) F_{2} + (a_{1} + ib_{1}) F_{3} + (a_{2} + ib_{2}) iF_{0} - (a_{3} + ib_{3}) F_{1} = E_{2} + i cB_{2} \; ,
\nonumber
\\
-(a_{0} + ib_{0}) F_{3} - (a_{1} + ib_{1}) F_{2} + (a_{2} + ib_{2}) F_{1} + (a_{3} + ib_{3})i F_{0} = E_{3} + i cB_{3} \; ,
\nonumber
\end{eqnarray}

\noindent from whence  it follows
\begin{eqnarray}
-b_{0} F_{0} + a_{1} F_{1} + a_{2}F_{2} +a_{3}F_{3} =0 \; , \qquad \qquad
a_{0} F_{0} + b_{1} F_{1} + b_{2}F_{2} +b_{3}F_{3} = 0\; ,
\nonumber
\\
-a_{0} F_{1} - b_{1} F_{0} - a_{2}F_{3} + a_{3}F_{2} = E_{1}\; , \qquad
-b_{0} F_{1} + a_{1} F_{0} - b_{2}F_{3} + b_{3}F_{2} = cB_{1}\; ,
\nonumber
\\
-a_{0}F_{2} + a_{1}F_{3} - b_{2} F_{0} - a_{3} F_{1} = E_{2} \; , \qquad
-b_{0} F_{2} + b_{1} F_{3} + a_{2}F_{0} - b_{3}F_{1} = cB_{2} \; ,
\nonumber
\\
-a_{0}F_{3} - a_{1}F_{2} + a_{2}F_{1} - b_{3}F_{0} = E_{3}\;, \qquad
-b_{0}F_{3} - b_{1}F_{2} + b_{2}F_{1} + a_{3}F_{0} = cB_{3} \; .
\label{5.3}
\end{eqnarray}

Let us consider  equation $ \mbox{rot}\; {\bf E} = - \partial_{0} c{\bf B}$.
Taking two identities
\begin{eqnarray}
\partial_{1} E_{2} - \partial_{2} E_{1}=
\partial_{1}
[ -a_{0}F_{2} + a_{1}F_{3} - b_{2} F_{0} - a_{3} F_{1} ] \;-
\nonumber
\\
- \partial_{2} [ -a_{0} F_{1} - b_{1} F_{0} - a_{2}F_{3} + a_{3}F_{2} ] \;=
\nonumber
\\
=  a_{1} \partial_{1} F_{3} - b_{2}  \partial_{1} F_{0} - a_{3}  \partial_{1} F_{1}  +
  b_{1}  \partial_{2} F_{0} + a_{2} \partial_{2} F_{3} - a_{3} \partial_{2} F_{2}  \; ,
\nonumber
\\
-\partial_{0} cB_{3} =  b_{0} \partial_{0} F_{3}  +  b_{1} \partial_{0} F_{2} - b_{2} \partial_{0} F_{1}
- a_{3} \partial_{0} F_{0} \; ;
\nonumber
\end{eqnarray}

\noindent
we produce an equation
\begin{eqnarray}
 a_{1} \partial_{1} F_{3} - b_{2}  \partial_{1} F_{0} - a_{3}  \partial_{1} F_{1}  +
  b_{1}  \partial_{2} F_{0} + a_{2} \partial_{2} F_{3} - a_{3} \partial_{2} F_{2}=
\nonumber
\\
=
b_{0} \partial_{0} F_{3}  +  b_{1} \partial_{0} F_{2} - b_{2} \partial_{0} F_{1}
- a_{3} \partial_{0} F_{0} \; ,
\nonumber
\end{eqnarray}

\noindent from whence
substituting identity $\partial_{0} F_{0} = \partial_{1} F_{1} + \partial_{2} F_{2} + \partial_{3} F_{3}$,
we obtain
\begin{eqnarray}
 a_{1} \partial_{1} F_{3} + a_{2} \partial_{2} F_{3} =
b_{0} \partial_{0} F_{3} - a_{3}     \partial_{3} F_{3} \; ;
\nonumber
\end{eqnarray}

\noindent
that is
\begin{eqnarray}
[\; b_{0}\partial_{0} - ( a_{1} \partial_{1} + a_{2} \partial_{2} + a_{3} \partial_{3} ) \;] \;  F_{3} = 0 \; .
\label{5.4}
\end{eqnarray}

\noindent
In the same manner we get
\begin{eqnarray}
[\; b_{0}\partial_{0} - ( a_{1} \partial_{1} + a_{2} \partial_{2} + a_{3} \partial_{3} ) \;] \;  F_{1} = 0 \; ,
\nonumber
\\
\; [\; b_{0}\partial_{0} - ( a_{1} \partial_{1} + a_{2} \partial_{2} + a_{3} \partial_{3} ) \;] \;  F_{2} = 0 \; .
\label{5.4'}
\end{eqnarray}

Now consider equation $\mbox{rot}\; c{\bf B} = \partial_{0} {\bf E}$. Calculating two rerms
\begin{eqnarray}
\partial_{1} cB_{2} -  \partial_{2} cB_{1} =
\nonumber
\\
= b_{1}  \partial_{1} F_{3} +
a_{2} \partial_{1} F_{0} - b_{3} \partial_{1}   F_{1}  - a_{1}  \partial_{2} F_{0} +
b_{2} \partial_{2} F_{3} - b_{3} \partial_{2} F_{2}\; ,
\nonumber
\\
\partial_{0} E_{3} = -a_{0} \partial_{0} F_{3} - a_{1} \partial_{0} F_{2} +
 a_{2} \partial_{0}F_{1} - b_{3} \partial_{0} F_{0} =
 \nonumber
 \\
 = -a_{0} \partial_{0} F_{3} - a_{1} \partial_{0} F_{2} +
 a_{2} \partial_{0}F_{1} - b_{3}  \partial_{1} F_{1} - b_{3}  \partial_{2} F_{2}
 -b_{3}  \partial_{3} F_{3} \; ,
 \nonumber
 \end{eqnarray}

\noindent
 we arrive at
\begin{eqnarray}
b_{1}  \partial_{1} F_{3} +
a_{2} \partial_{1} F_{0} - b_{3} \partial_{1}  F_{1}  - a_{1}  \partial_{2} F_{0} +
b_{2} \partial_{2} F_{3} - b_{3} \partial_{2} F_{2}  =
\nonumber
\\
= -a_{0} \partial_{0} F_{3} - a_{1} \partial_{0} F_{2} +
 a_{2} \partial_{0}F_{1} - b_{3}  \partial_{1} F_{1} - b_{3}  \partial_{2} F_{2}
 -b_{3}  \partial_{3} F_{3} \; ,
\nonumber
\end{eqnarray}

\noindent or
\begin{eqnarray}
[ \; a_{0} \partial_{0}      + ( b_{1}  \partial_{1}  +
b_{2} \partial_{2} + b_{3}  \partial_{3}  ) \; ]  \; F_{3}  = 0 \; .
\label{5.5}
\end{eqnarray}

\noindent
Analogously, we get
\begin{eqnarray}
[ \; a_{0} \partial_{0}      + ( b_{1}  \partial_{1}  +
b_{2} \partial_{2} + b_{3}  \partial_{3}  ) \; ]  \; F_{1}  = 0 \; ,
\nonumber
\\
\; [ \; a_{0} \partial_{0}      + ( b_{1}  \partial_{1}  +
b_{2} \partial_{2} + b_{3}  \partial_{3}  ) \; ]  \; F_{2}  = 0 \; .
\label{5.5'}
\end{eqnarray}

Now let us consider equation $\mbox{div}\; {\bf E} = 0$:
\begin{eqnarray}
0 = \partial_{1} E_{1} + \partial_{2} E_{2} +  \partial_{3} E_{3} =
-a_{0} \partial_{1}  F_{1} - b_{1} \partial_{1}  F_{0} -
a_{2} \partial_{1} F_{3} + a_{3} \partial_{1} F_{2}  -
\nonumber
\\
 -a_{0} \partial_{2} F_{2} + a_{1} \partial_{2} F_{3} -
b_{2}  \partial_{2} F_{0} - a_{3} \partial_{2}  F_{1}   -
   a_{0} \partial_{3} F_{3} - a_{1} \partial_{3} F_{2} +
  a_{2} \partial_{3} F_{1} - b_{3} \partial_{3} F_{0} \; ,
\nonumber
\end{eqnarray}

\noindent that is
\begin{eqnarray}
-a_{0} (\partial_{1} F_{1} + \partial_{2} F_{2} + \partial_{3} F_{3}) -
(b_{1} \partial_{1} + b_{2} \partial_{2} + b_{3} \partial_{3}) F_{0}  = 0 \; ,
\end{eqnarray}

\noindent
which  is equivalent to
\begin{eqnarray}
[ \; a_{0} \partial_{0}      + ( b_{1}  \partial_{1}  +
b_{2} \partial_{2} + b_{3}  \partial_{3}  )\;  ]  \; F_{0}  = 0 \; .
\label{5.6}
\end{eqnarray}

It remains to consider equation $\mbox{div}\; {\bf B} = 0$:
\begin{eqnarray}
0= -b_{0} \partial_{1}  F_{1} + a_{1} \partial_{1}  F_{0} -
b_{2} \partial_{1} F_{3} + b_{3} \partial_{1} F_{2} -
b_{0}  \partial_{2} F_{2} + b_{1}\partial_{2}  F_{3} +
a_{2} \partial_{2} F_{0} - b_{3} \partial_{2} F_{1} -
\nonumber
\\
-b_{0} \partial_{3}F_{3} - b_{1} \partial_{3} F_{2} +
b_{2} \partial_{3} F_{1} + a_{3} \partial_{3} F_{0}  \; ,
\nonumber
\end{eqnarray}

\noindent  or
\begin{eqnarray}
-b_{0} (\partial_{1} F_{1} + \partial_{2} F_{2} + \partial_{3} F_{3}) +
(a_{1} \partial_{1} + a_{2} \partial_{2} + a_{3} \partial_{3}) F_{0} = 0 \; ,
\nonumber
\end{eqnarray}

\noindent
which  is equivalent to
\begin{eqnarray}
[ b_{0} \partial_{0}      - ( a_{1}  \partial_{1}  +
a_{2} \partial_{2} + a_{3}  \partial_{3}  ) ]  \; F_{0}  = 0 \; .
\label{5.7}
\end{eqnarray}

Thus, to construct physical solutions of the Maxwell equations as linear combinations from non-physical ones
  (see (\ref{5.2}) )
\begin{eqnarray}
(a_{0} + i b_{0})  \Psi^{0}  +  (a_{1} + i b_{1}) \Psi^{1} +
(a_{2} + i b_{2}) \Psi^{2} +  (a_{3} + i b_{3}) \Psi^{3}  =
\left |
\begin{array}{c} 0 \\ {\bf E} + i c{\bf B}
\end{array} \right |
\label{5.8}
\end{eqnarray}

\noindent  one must satisfy  the following 8 equations:
\begin{eqnarray}
 [\; b_{0}\partial_{0} - ( a_{1} \partial_{1} + a_{2} \partial_{2} + a_{3}
 \partial_{3} ) \;] \;  F_{c} = 0 \; ,
\nonumber
\\[2mm]
\; [ \; a_{0} \partial_{0}      + ( b_{1}  \partial_{1}  +
b_{2} \partial_{2} + b_{3}  \partial_{3}  ) \; ]  \; F_{c}  = 0 \; ; \;
\label{5.9}
\end{eqnarray}

\noindent where $  c = 0,1,2,3 ; \; F_{c} = \partial_{c} \Phi  $.

\section{  On  separating physical solutions of the  Maxwell  equations \\ (complex-valued scalar function $\Phi $ ) }

Let us consider four  formal solutions of the Maxwell equations
\begin{eqnarray}
\{ \Psi^{0} ,  \Psi^{1}, \Psi^{2} ,\Psi^{3} \} =
\left | \begin{array}{rrrr}
i F_{0}  &  \quad  F_{1} &  \quad  F_{2}  &  \quad  F_{3}   \\
 -F_{1}  &  \quad iF_{0} &  \quad -F_{3}  &  \quad  F_{2}  \\
 -F_{2}  &  \quad  F_{3} &  \quad iF_{0}  &  \quad -F_{1}   \\
 -F_{3}  &  \quad -F_{2}  &  \quad F_{1}  &  \quad iF_{0}
\end{array}  \right | \; , \qquad  F_{a} (x) = \partial_{a} \Phi (x) \; .
\label{A.1}
\end{eqnarray}

Let the function  $\Phi (x)$ be  taken as  complex-valued.
We should examine relationship for $\lambda_{a}$ defining all  possible solutions to the following equation:
\begin{eqnarray}
\lambda _{0} \Psi^{0}  +  \lambda _{1} \Psi^{1} + \lambda _{2} \Psi^{2} + \lambda _{3}\Psi^{3}  =
\left |
\begin{array}{c} 0 \\ {\bf E} + i c{\bf B}
\end{array} \right |.
\label{A.2}
\end{eqnarray}

\noindent
Let us separate real and imaginary parts in $\lambda_{c}$ and $\Phi (x)$ and $F_{c}(x)$:
\begin{eqnarray}
\lambda_{c} = a_{c} + i \; b_{c} \; , \qquad \Phi (x) = L (x) + i \; K(x) \;,
\qquad
F_{c} (x) = L _{c}(x) + i \; K_{c}(x) \; .
\nonumber
\end{eqnarray}

\noindent
 Relation
(\ref{A.2}) takes the form
\begin{eqnarray}
(a_{0} + ib_{0}) i (L_{0} + i K_{0}) + (a_{1} + ib_{1}) (L_{1} + i K_{1}) +
\nonumber
\\
 +
(a_{2} + ib_{2}) (L_{2} + i K_{2}) + (a_{3} + ib_{3}) (L_{3} + i K_{3}) = 0 \; ,
\nonumber
\\[3mm]
-(a_{0} + ib_{0}) (L_{1} + i K_{1}) + (a_{1} + ib_{1}) i (L_{0} + i K_{0}) -
\nonumber
\\
-
(a_{2} + ib_{2}) (L_{3} + i K_{3}) + (a_{3} + ib_{3}) (L_{2} + i K_{2}) = E_{1} + i cB_{1} \; ,
\nonumber
\\[3mm]
-(a_{0} + ib_{0}) (L_{2} + i K_{2}) + (a_{1} + ib_{1}) (L_{3} + i K_{3}) +
\nonumber
\\
+ (a_{2} + ib_{2}) i (L_{0} + i K_{0})  - (a_{3} + ib_{3}) (L_{1} + i K_{1}) = E_{2} + i cB_{2} \; ,
\nonumber
\\[3mm]
-(a_{0} + ib_{0}) (L_{3} + i K_{3}) - (a_{1} + ib_{1}) (L_{2} + i K_{2}) +
\nonumber
\\
+ (a_{2} + ib_{2}) (L_{1} + i K_{1}) + (a_{3} + ib_{3})i (L_{0} + i K_{0}) = E_{3} + i cB_{3} \; ,
\nonumber
\end{eqnarray}

\noindent from whence  it follows
\begin{eqnarray}
0 = (-b_{0} L_{0} + a_{1} L_{1} + a_{2}L_{2} +a_{3}L_{3}) + (-a_{0} K_{0} - b_{1} K_{1} - b_{2}K_{2} -b_{3}K_{3} ) \; ,
\nonumber
\\
0= (+ a_{0} L_{0} + b_{1} L_{1} + b_{2}L_{2} +b_{3}L_{3}) + (-b_{0} K_{0} +a_{1} K_{1} + a_{2}K_{2} + a_{3}K_{3} ) \; ,
\nonumber
\\[3mm]
 E_{1} = (-a_{0} L_{1} - b_{1} L_{0} - a_{2} L_{3 } +a_{3} L_{2}) +
(+b_{0} K_{1} - a_{1} K_{0} + b_{2}K_{3} -b_{3}K_{2} ) \; ,
\nonumber
\\
B_{1} =
(-b_{0} L_{1} + a_{1} L_{0} - b_{2} L_{3 } + b_{3} L_{2}) +
(-a_{0} K_{1} - b_{1} K_{0} - a_{2} K_{3}  + a_{3} K_{2} ) \; ,
\nonumber
\\[3mm]
 E_{2} = ( - a_{0} L_{2} - b_{2} L_{0} + a_{1} L_{3 } - a_{3} L_{1} ) +
         ( + b_{0} K_{2} - a_{2} K_{0} - b_{1} K_{3}  + b_{3} K_{1} ) \; ,
\nonumber
\\
B_{2} =
      ( - b_{0} L_{2 } + a_{2} L_{0} + b_{1} L_{3 } - b_{3} L_{1}) +
      ( - a_{0} K_{2} - b_{2}  K_{0} + a_{1} K_{3}  - a_{3} K_{1} ) \; ,
\nonumber
\\[3mm]
 E_{3} = ( - a_{0} L_{3} - b_{3} L_{0} - a_{1} L_{2 } + a_{2} L_{1} ) +
         ( + b_{0} K_{3} - a_{3} K_{0} + b_{1} K_{2}  - b_{2} K_{1} ) \; ,
\nonumber
\\
B_{3} =
      ( - b_{0} L_{3 } + a_{3} L_{0} - b_{1} L_{2 } + b_{2} L_{1}) +
      ( - a_{0} K_{3} - b_{3}  K_{0} - a_{1} K_{2}  + a_{2} K_{1} ) \; ,
\label{A.3}
\end{eqnarray}

Substituting these expression   into Maxwell equations and performing calculation like in previous section  we get:

\begin{eqnarray}
\mbox{div} \; {\bf E} = 0 \qquad \Longrightarrow \qquad
-(a_{0} \partial_{0} + b_{j}\partial_{j} )\; L_{0} +
(b_{0} \partial_{0} - a_{j}\partial_{j} )\; K_{0} = 0 \; ,
\nonumber
\\
\mbox{div} \; {\bf B} = 0 \qquad \Longrightarrow \qquad
+(b_{0} \partial_{0} - a_{j}\partial_{j} )\; L_{0} +
(a_{0} \partial_{0} + b_{j}\partial_{j} )\; K_{0} = 0 \; ,
\nonumber
\\[3mm]
\partial_{2}B_{3} - \partial_{3} B_{2} = + \partial_{0} E_{1}
\qquad \Longrightarrow \qquad
-(a_{0} \partial_{0} + b_{j}\partial_{j} )\; L_{1} +
(b_{0} \partial_{0} - a_{j}\partial_{j} )\; K_{1} = 0 \; ,
\nonumber
\\
\partial_{2}E_{3} - \partial_{3} E_{2} = - \partial_{0}B_{1}
\qquad \Longrightarrow \qquad
+(b_{0} \partial_{0} - a_{j}\partial_{j} )\; L_{1} +
(a_{0} \partial_{0} + b_{j}\partial_{j} )\; K_{1} = 0 \; ,
\nonumber
\\[3mm]
\partial_{3}B_{1} - \partial_{1} B_{3} = + \partial_{0} E_{2}
\qquad \Longrightarrow \qquad
-(a_{0} \partial_{0} + b_{j}\partial_{j} )\; L_{2} +
(b_{0} \partial_{0} - a_{j}\partial_{j} )\; K_{2} = 0 \; ,
\nonumber
\\
\partial_{3}E_{1} - \partial_{1} E_{3} = - \partial_{0}B_{2}
\qquad \Longrightarrow \qquad
+(b_{0} \partial_{0} - a_{j}\partial_{j} )\; L_{2} +
(a_{0} \partial_{0} + b_{j}\partial_{j} )\; K_{2} = 0 \; ,
\nonumber
\\[3mm]
\partial_{1}B_{2} - \partial_{2} B_{1} = + \partial_{0} E_{3}
\qquad \Longrightarrow \qquad
-(a_{0} \partial_{0} + b_{j}\partial_{j} )\; L_{3} +
(b_{0} \partial_{0} - a_{j}\partial_{j} )\; K_{3} = 0 \; ,
\nonumber
\\
\partial_{1}E_{2} - \partial_{2} E_{1} = - \partial_{0}B_{3}
\qquad \Longrightarrow \qquad
+(b_{0} \partial_{0} - a_{j}\partial_{j} )\; L_{3} +
(a_{0} \partial_{0} + b_{j}\partial_{j} )\; K_{3} = 0 \; ,
\end{eqnarray}

Thus, for physical Maxwell solutions the following equations must hold:
\begin{eqnarray}
-(a_{0} \partial_{0} + b_{j}\partial_{j} )\; L_{c} +
(b_{0} \partial_{0} - a_{j}\partial_{j} )\; K_{c} = 0 \; ;
\nonumber
\\
+(b_{0} \partial_{0} - a_{j}\partial_{j} )\; L_{c} +
(a_{0} \partial_{0} + b_{j}\partial_{j} )\; K_{c} = 0 \; .
\label{A4}
\end{eqnarray}

In particular, we have two more simple  equations when taking real or imaginary  scalar functions:

\vspace{5mm}

$
\Phi = L+ i \; 0 \; ,
$
\begin{eqnarray}
 a_{c}, b_{c} \; \; \Longleftarrow \qquad  (a_{0} \; \partial_{0} + b_{j} \; \partial_{j} )\; L_{c} =0 \; , \;\;
(b_{0}  \; \partial_{0} - a_{j} \; \partial_{j} )\; L_{c} = 0
\nonumber
\end{eqnarray}

$\Phi = 0+ i \; K \;$
\begin{eqnarray}
a_{c}', b_{c}' \; \; \Longleftarrow \qquad   (a_{0}'  \; \partial_{0} + b_{j}' \; \partial_{j} )\; K_{c} =0 \; , \;\;
(b_{0}' \; \partial_{0} - a_{j}' \; \partial_{j} )\; K_{c} = 0 \; .
\nonumber
\\
\label{A5}
\end{eqnarray}

In the simplest case of a plane scalar wave
\begin{eqnarray}
\Phi = e^{i (k_{0} x^{0} -  k^{j} x^{j} )} = \cos  \varphi  + i
\sin  \varphi \; ,
\nonumber
\\
 \varphi = k_{0} \;  x^{0} -  k^{j} \; x^{j}  = k_{c}  \; x^{c}  \; ,
\\
\nonumber
\qquad L_{c} + i K_{c} = - k_{c} \sin  \varphi  + i \;
k_{c} \cos  \varphi \nonumber
\end{eqnarray}

\noindent
previous equations  take the form

\vspace{5mm}

$
\Phi = L+ i \; 0 \; ,
$
\begin{eqnarray}
 a_{c}, b_{c} \; \; \Longleftarrow \qquad  (a_{0} \; k_{0} + b_{j} \; k_{j} )\; k_{c}    =0 \; , \;\;
(b_{0}  \; k_{0} - a_{j} \; k_{j} )\; k_{c}   = 0
\nonumber
\end{eqnarray}

$\Phi = 0+ i \; K \;$
\begin{eqnarray}
a_{c}', b_{c}' \; \; \Longleftarrow \qquad   (a_{0}'  \; k_{0} + b_{j}' \; k_{j} )\; k_{c} =0 \; , \;\;
(b_{0}' \; k_{0} - a_{j}' \; k_{j} )\; k_{c} = 0 \; .
\nonumber
\\
\label{A6}
\end{eqnarray}

that is
\begin{eqnarray}
a'_{c} = a_{c} \; , \qquad b'_{c} = b_{c} \; ,
\nonumber
\\
(a_{0} \; k_{0} + b_{j} \; k_{j} )\; k_{c}    =0 \; , \;\;
(b_{0}  \; k_{0} - a_{j} \; k_{j} )\; k_{c}   = 0
\label{A7}
\end{eqnarray}

In general case ({\ref{A4}) equations for  $a_{c}, b_{c}$ and   $a_{c}', b_{c}'$  may not coincide.

Let us turn again to  eqs. ({\ref{A5}) and translate them to variables
\begin{eqnarray}
L_{c} = { F_{c} ^{*} + F_{c} \over 2} \; , \qquad
K_{c} =  +i \; { F_{c}^{*} - F_{c} \over 2} \; , \qquad
\nonumber
\end{eqnarray}

\begin{eqnarray}
-(a_{0} \partial_{0} + b_{j}\partial_{j} )\;  { F_{c} ^{*} + F_{c} \over 2}  +
(b_{0} \partial_{0} - a_{j}\partial_{j} )\; i \; { F_{c}^{*} - F_{c} \over 2}  = 0 \; ;
\nonumber
\\
+(b_{0} \partial_{0} - a_{j}\partial_{j} )\; { F_{c} ^{*} + F_{c} \over 2}  +
(a_{0} \partial_{0} + b_{j}\partial_{j} )\; i \; { F_{c}^{*} - F_{c} \over 2} = 0 \; .
\nonumber
\end{eqnarray}

or
\begin{eqnarray}
{1 \over 2} \; [ \; -(a_{0} \partial_{0} + b_{j}\partial_{j} )  - i \;  (b_{0} \partial_{0} - a_{j}\partial_{j} ) \;  ] \;  F_{c} +
{1 \over 2} \; [  \; - (a_{0} \partial_{0} + b_{j}\partial_{j} ) + i \;  (b_{0} \partial_{0} - a_{j}\partial_{j} )\;  ] \; F^{*}_{c} = 0
\nonumber
\\
{1 \over 2} \; [ \; + (b_{0} \partial_{0} - a_{j}\partial_{j} )  - i\;
(a_{0} \partial_{0} + b_{j}\partial_{j} ) \; ] \; F_{c} +
{1 \over 2} \;
[ \;+ (b_{0} \partial_{0} - a_{j}\partial_{j} )  + i\;
(a_{0} \partial_{0} + b_{j}\partial_{j} ) \; ] \; F_{c}^{*} =0 \; ;
\nonumber
\end{eqnarray}

\noindent or

\begin{eqnarray}
{1 \over 2} \; [ \; -(a_{0} \partial_{0} + b_{j}\partial_{j} )  - i \;  (b_{0} \partial_{0} - a_{j}\partial_{j} ) \;  ] \;  F_{c} +
{1 \over 2} \; [  \; - (a_{0} \partial_{0} + b_{j}\partial_{j} ) + i \;  (b_{0} \partial_{0} - a_{j}\partial_{j} )\;  ] \; F^{*}_{c} = 0
\nonumber
\\
{1 \over 2} \; [ \; + i \; (b_{0} \partial_{0} - a_{j}\partial_{j} )  +
(a_{0} \partial_{0} + b_{j}\partial_{j} ) \; ] \; F_{c} +
{1 \over 2} \;
[ \;+ i\; (b_{0} \partial_{0} - a_{j}\partial_{j} )  -
(a_{0} \partial_{0} + b_{j}\partial_{j} ) \; ] \; F_{c}^{*} =0 \; ;
\nonumber
\end{eqnarray}

Summing and subtracting two relations, we arrive at
\begin{eqnarray}
[  \; - (a_{0} \partial_{0} + b_{j}\partial_{j} ) +
i \;  (b_{0} \partial_{0} - a_{j}\partial_{j} )\;  ] \; F^{*}_{c} = 0
\nonumber
\\
\; [ \; -(a_{0} \partial_{0} + b_{j}\partial_{j} )  - i \;  (b_{0} \partial_{0} - a_{j}\partial_{j} ) \;  ] \;  F_{c} = 0
\label{A8}
\end{eqnarray}

They can be rewritten as follows:
\begin{eqnarray}
[\;  - \; (a_{0} + i b_{0} ) \; \partial _{0} + i ( a_{j} + i b_{j} ) \; \partial_{j} \; ] \; F_{c} = 0 \; ,
\nonumber
\\
\;
[\;  - \; (a_{0} - i b_{0} ) \; \partial _{0} - i ( a_{j} - i b_{j} ) \; \partial_{j} \; ] \; F_{c} ^{*}= 0 \; ,
\nonumber
\end{eqnarray}

or shorter

\begin{eqnarray}
[\;  - \; \lambda_{0} \; \partial _{0} + i \; \lambda_{j}  \; \partial_{j} \; ] \; F_{c} = 0 \; ,
\nonumber
\\
\;
[\;  - \; \lambda_{0} \; \partial _{0}^{*} - i  \; \lambda_{j}^{*} \; \partial_{j} \; ] \; F_{c} ^{*}= 0 \; ,
\label{A9}
\end{eqnarray}

\section{ Separating physical solutions of the plane wave type}

Let us apply  the general relations (\ref{5.9})  to the  case when
\begin{eqnarray}
\Phi =A  \; \sin \varphi \; , \qquad \varphi =   k_{0} x_{0} - k_{3} x_{3} \; ,
\nonumber
\\
  F_{0} = k_{0}A \cos \varphi  \; , \;\; F_{1} = 0 \; , \;\; F_{2} = 0 \; , \;
  F_{3} = -k_{3}A \cos \varphi \;  .
\nonumber
\end{eqnarray}

\noindent
Eqs. (\ref{5.9}) then give
\begin{eqnarray}
b_{0} k_{0} + a_{3} k_{3} = 0 \; , \qquad
a_{0} k_{0} - b_{3} k_{3} = 0  \; .
\label{5.10}
\end{eqnarray}

For a wave spreading in the positive direction
 $k_{3} = + k_{0} > 0 $, and eqs. (\ref{5.10}) give
\begin{eqnarray}
b_{0} =- a_{3}   \; , \qquad  b_{3} = a_{0}    \; ,
\nonumber
\end{eqnarray}

\noindent and correspondingly relationship  (\ref{5.8}) looks
\begin{eqnarray}
 (a_{0} - i a_{3})  \Psi^{0}   +  (a_{3} + i a_{0}) \Psi^{3}  +  (a_{1} + i b_{1}) \Psi^{1} +
(a_{2} + i b_{2}) \Psi^{2}     =
\left |
\begin{array}{c} 0 \\ {\bf E} + i c{\bf B}
\end{array} \right | \; ;
\label{5.11}
\end{eqnarray}

\noindent coefficients at $\Psi^{1}$ and $\Psi^{2}$ are arbitrary.
To understand this fact let us recall the explicit form of $\Psi^{a}$ --
see (\ref{3.7}):
\begin{eqnarray}
\{ \Psi^{0} ,  \Psi^{1}, \Psi^{2}, \Psi^{3} \}  =  A \left | \begin{array}{rrrr}
i k_{0} & 0  &  0  & - k_{0}  \\
0 & i k_{0}  &  k_{0}  & 0  \\
0 & - k_{0}  &  i k_{0}  & 0  \\
 k_{0} & 0  &  0  & i k_{0}
\end{array}  \right |  \cos \varphi \; .
\nonumber
\end{eqnarray}

\noindent
One may separate two subsets of non-physical solutions :
\begin{eqnarray}
(a_{0} - i a_{3})  \Psi^{0}   +  (a_{3} + i a_{0}) \Psi^{3}  =
(a_{0} - i a_{3}) k_{0} \;  \left \{ \;\;\;
\left | \begin{array}{r}
 i \\ 0 \\ 0 \\  +1
\end{array} \right | +
  i \;
  \left | \begin{array}{r}
    - 1 \\ 0 \\ 0  \\  i
\end{array} \right | \;\;\; \right \} \equiv
\left | \begin{array}{c}
0 \\ 0  \\ 0  \\ 0
\end{array} \right | \; ,
\label{5.12}
\end{eqnarray}

\noindent
so relationship (\ref{5.11}) reduces to
\begin{eqnarray}
(a_{1} + i b_{1})  \Psi^{1}   +  (a_{2} + i b_{2}) \Psi^{2}  =
(a_{1} + i b_{1})
\left | \begin{array}{r}
0 \\ ik_{0} \\ -k_{3} \\ 0
\end{array} \right | +
  (a_{2} + i b_{2})
  \left | \begin{array}{r}
    0 \\ k_{3} \\ ik_{3} \\ 0
\end{array} \right |\; .
\label{5.12}
\end{eqnarray}

Now let us consider a more general case  when
\begin{eqnarray}
\Phi (x) = A \sin ( k_{0}x_{0} - {\bf k} {\bf x} ) = A \;  \sin \varphi \; ,
\nonumber
\\
 F_{0} =    k_{0} A  \cos \varphi \; , \qquad
F_{1} =  -   k_{1} A \cos \varphi  \; ,
\nonumber
\\
F_{2} = -   k_{2} A \cos \varphi  \; , \qquad
 F_{3} =  -   k_{3} A \cos \varphi  \;  .
\label{5.13}
\end{eqnarray}

\noindent Eqs. (\ref{5.9}) give
\begin{eqnarray}
  b_{0} k_{0} +  a_{1} k_{1} + a_{2} k_{2} + a_{3}
 k_{3}   = 0 \; ,
\qquad
 a_{0} k_{0}      - b_{1}  k_{1}  - b_{2} k_{2} - b_{3}  k_{3}    = 0 \; ,
\label{5.14}
\end{eqnarray}

\noindent and additionally the identity
$
k_{0} = + \sqrt{ k_{1}^{2} + k_{2}^{2} + k_{3}^{2} } \;
$
 holds. One may introduce parametrization $k_{j} = k_{0} n_{j}, \; n_{j} n_{j} = 1 $, then
eqs. (\ref{5.14}) read
\begin{eqnarray}
  b_{0}  = - ( a_{1} n_{1} + a_{2} n_{2} + a_{3}n_{3} )   \; , \qquad
 a_{0}   =  (b_{1}  n_{1}  + b_{2} n_{2} + b_{3}  n_{3} )   \; .
\label{5.15}
\end{eqnarray}

Now turning to (\ref{5.8})
 and excluding variables $a_{0}, b_{0}$ one gets
\begin{eqnarray}
[(b_{1}  n_{1}  + b_{2} n_{2} + b_{3}  n_{3} ) - i ( a_{1} n_{1} + a_{2} n_{2} + a_{3}n_{3} )  ] \Psi^{0}  +
\nonumber
\\
+  (a_{1} + i b_{1}) \Psi^{1} +
(a_{2} + i b_{2}) \Psi^{2} +  (a_{3} + i b_{3}) \Psi^{3}  = \Psi  \; .
\label{5.16}
\end{eqnarray}

\noindent Taking into account (see (\ref{5.1})
\begin{eqnarray}
\{ \Psi^{0} ,  \Psi^{1}, \Psi^{2} ,\Psi^{3} \} =
\left | \begin{array}{rrrr}
i          &  \quad  -n_{1} &  \quad  -n_{2}  &  \quad  -n_{3}   \\
 n_{1}  &  \quad i &  \quad n_{3}  &  \quad  -n_{2}  \\
 n_{2}  &  \quad  -n_{3} &  \quad i  &  \quad n_{1}   \\
 n_{3}  &  \quad n_{2}  &  \quad -n_{1}  &  \quad i
\end{array}  \right |  \; k_{0} A \cos \varphi   \; ,
\nonumber
\end{eqnarray}

\noindent
from (\ref{5.16}) we get (factor $ k_{0} A \cos \varphi $ is omitted)
 \begin{eqnarray}
\Psi =  \left | \begin{array}{r}
\;[ (b_{1}-i a_{1}) n_{1}  + (b_{2}-i a_{2}) n_{2} + (b_{3}-i a_{3}) n_{3} ] \; i    - (a_{1} + i b_{1}) n_{1}  - (a_{2} + i b_{2})n_{2}    - (a_{3} + i b_{3})n_{3}   \\
\;[(b_{1}-i a_{1}) n_{1}  + (b_{2}-i a_{2}) n_{2} + (b_{3}-i a_{3}) n_{3} ] \; n_{1}+ (a_{1} + i b_{1}) i     + (a_{2} + i b_{2})   n_{3}  - (a_{3} + i b_{3}) n_{2}  \\
\;[(b_{1}-i a_{1}) n_{1}  + (b_{2}-i a_{2}) n_{2} + (b_{3}-i a_{3}) n_{3} ] \; n_{2}- (a_{1} + i b_{1}) n_{3} + (a_{2} + i b_{2}) i    +(a_{3} + i b_{3}) n_{1}   \\
\;[(b_{1}-i a_{1}) n_{1}  + (b_{2}-i a_{2}) n_{2} + (b_{3}-i a_{3}) n_{3} ] \; n_{3}+ (a_{1} + i b_{1}) n_{2} - (a_{2} + i b_{2})n_{1} + (a_{3} + i b_{3}) i
\end{array}  \right |
\nonumber
\end{eqnarray}

\noindent and further

 \begin{eqnarray}
\Psi =  \left | \begin{array}{c}
0 \\
\;[(b_{1}-i a_{1}) n_{1}  + (b_{2}-i a_{2}) n_{2} + (b_{3}-i a_{3}) n_{3} ] \; n_{1}+ (a_{1} + i b_{1}) i     + (a_{2} + i b_{2})   n_{3}  - (a_{3} + i b_{3}) n_{2}  \\
\;[(b_{1}-i a_{1}) n_{1}  + (b_{2}-i a_{2}) n_{2} + (b_{3}-i a_{3}) n_{3} ] \; n_{2}- (a_{1} + i b_{1}) n_{3} + (a_{2} + i b_{2}) i    +(a_{3} + i b_{3}) n_{1}   \\
\;[(b_{1}-i a_{1}) n_{1}  + (b_{2}-i a_{2}) n_{2} + (b_{3}-i a_{3}) n_{3} ] \; n_{3}+ (a_{1} + i b_{1}) n_{2} - (a_{2} + i b_{2})n_{1} + (a_{3} + i b_{3}) i
\end{array}  \right |    .
\nonumber\\
\label{5.17}
\end{eqnarray}

\noindent
Let us introduce three elementary  solutions, associated with
coefficients $ (a_{j}+ib_{j})$:
 \begin{eqnarray}
\Psi_{(1)} =  \left | \begin{array}{l}
0 \\
(b_{1}-i a_{1})\;  n_{1}   n_{1} + (a_{1} + i b_{1}) i     \\
(b_{1}-i a_{1}) \; n_{1}   n_{2} - (a_{1} + i b_{1}) n_{3}  \\
(b_{1}-i a_{1})\;  n_{1}   n_{3} + (a_{1} + i b_{1}) n_{2}
\end{array}  \right |    ,
{\bf E} _{(1)} =
\left | \begin{array}{l}
b_{1}  n_{1}   n_{1} -  b_{1}     \\
b_{1} n_{1}   n_{2} - a_{1} n_{3}  \\
b_{1} n_{1}   n_{3} + a_{1} n_{2}
\end{array}  \right |    ,
c {\bf B} _{(1)} =  \left | \begin{array}{l}
- a_{1}   n_{1}   n_{1} + a_{1}     \\
- a_{1}  n_{1}   n_{2} -  b_{1} n_{3}  \\
- a_{1}  n_{1}   n_{3} +  b_{1} n_{2}
\end{array}  \right |    ,
\nonumber
\\[2mm]
\Psi_{(2)} =  \left | \begin{array}{l}
0 \\
(b_{2}-i a_{2})\;  n_{2}   n_{1}  + (a_{2} + i b_{2}) n_{3}   \\
(b_{2}-i a_{2})\;  n_{2}   n_{2}  + (a_{2} + i b_{2}) i    \\
(b_{2}-i a_{2}) \; n_{2}   n_{3}  - (a_{2} + i b_{2}) n_{1}
\end{array}  \right |    ,
{\bf E}_{(2)} =
\left | \begin{array}{l}
b_{2}  n_{2}   n_{1}  + a_{2}  n_{3}   \\
b_{2}  n_{2}   n_{2}  -  b_{2}    \\
b_{2}  n_{2}   n_{3}  - a_{2}  n_{1}
\end{array}  \right |    ,
c {\bf B}_{(2)} =  \left | \begin{array}{l}
- a_{2}  n_{2}   n_{1}  + b_{2}   n_{3}   \\
- a_{2}  n_{2}   n_{2}  + a_{2}    \\
- a_{2}  n_{2}   n_{3}  - b_{2} n_{1}
\end{array}  \right |    ,
\nonumber
\\[2mm]
\Psi _{(3)} =  \left | \begin{array}{l}
0 \\
(b_{3}-i a_{3}) \; n_{3}   n_{1} - (a_{3} + i b_{3}) n_{2}  \\
(b_{3}-i a_{3}) \; n_{3}    n_{2} + (a_{3} + i b_{3}) n_{1}  \\
(b_{3}-i a_{3}) \; n_{3} \;   n_{3} + (a_{3} + i b_{3}) i
\end{array}  \right |,
{\bf E}_{(3)} =
\left | \begin{array}{l}
b_{3} n_{3}   n_{1} - a_{3} n_{2}  \\
b_{3} n_{3}    n_{2} + a_{3}  n_{1}  \\
b_{3} n_{3} \;   n_{3} -  b_{3}
\end{array}  \right | ,
c{\bf B}_{(3)} =  \left | \begin{array}{l}
- a_{3} n_{3}   n_{1} - b_{3} n_{2}  \\
- a_{3} n_{3}    n_{2} +  b_{3} n_{1}  \\
- a_{3} n_{3} \;   n_{3} + a_{3}
\end{array}  \right |     .
\nonumber
\\
\label{5.18}
\end{eqnarray}

Let us show that three types of solutions are linearly dependent.  It suffices to
examine their  linear combinations
(for definite consider electric field):
\begin{eqnarray}
A_{1}  \; {\bf E}_{(1)} + A_{2}  \; {\bf E}_{(2)} + A_{3}  \; {\bf E}_{(3)} = 0 \; ,
\nonumber
\end{eqnarray}

\noindent that is
\begin{eqnarray}
A_{1}   \left | \begin{array}{l}
b_{1}  n_{1}   n_{1} -  b_{1}     \\
b_{1} n_{1}   n_{2} - a_{1} n_{3}  \\
b_{1} n_{1}   n_{3} + a_{1} n_{2}
\end{array}  \right | +
A_{2}  \left | \begin{array}{l}
b_{2}  n_{2}   n_{1}  + a_{2}  n_{3}   \\
b_{2}  n_{2}   n_{2}  -  b_{2}    \\
b_{2}  n_{2}   n_{3}  - a_{2}  n_{1}
\end{array}  \right |  +
A_{3} \left | \begin{array}{l}
b_{3}  n_{3}   n_{1}  - a_{3}  n_{2} \\
b_{3} n_{3}    n_{2} + a_{3}  n_{1}  \\
b_{3} n_{3} \;   n_{3} -  b_{3}
\end{array}  \right | = \left | \begin{array}{c} 0 \\ 0  \\ 0  \end{array} \right | \; .
\nonumber
\end{eqnarray}

It remains to show that the determinant of the $3 \times 3$  matrix vanishes
\begin{eqnarray}
\mbox{det}  \; \left | \begin{array}{lll}
b_{1}  n_{1}n_{1} -  b_{1}      &  b_{2}  n_{2}   n_{1}  + a_{2}n_{3} & b_{3} n_{3}   n_{1}  - a_{3}  n_{2}  \\
b_{1}  n_{1}n_{2} - a_{1} n_{3} &  b_{2}  n_{2}   n_{2}  - b_{2}      & b_{3} n_{3}    n_{2} + a_{3}  n_{1}  \\
b_{1} n_{1}n_{3} + a_{1} n_{2}  &  b_{2}  n_{2}   n_{3}  - a_{2}n_{1} & b_{3} n_{3} \;   n_{3} -  b_{3}
\end{array} \right | = 0 \; .
\label{5.19}
\end{eqnarray}

\noindent
Let $b_{1}n_{1} =s_{1},  b_{2}n_{2} =s_{2}, b_{3}n_{3} =s_{3}$ then eq. (\ref{5.19}) looks
\begin{eqnarray}
0 = \left | \begin{array}{lll}
s_{1} n_{1} -  b_{1}       & \quad  s_{2}    n_{1}  + a_{2}n_{3}   & \quad  s_{3}   n_{1}  - a_{3}  n_{2}  \\
s_{1} n_{2} - a_{1} n_{3}  & \quad  s_{2}    n_{2}  - b_{2}        & \quad  s_{3}   n_{2} + a_{3}  n_{1}  \\
s_{1} n_{3} + a_{1} n_{2}  & \quad  s_{2}    n_{3}  - a_{2}n_{1}   &  \quad  s_{3}   n_{3} -  b_{3}
\end{array} \right | =
\nonumber
\\[3mm]
=
(s_{1} n_{1} -  b_{1})
[  (s_{2}    n_{2}  - b_{2} ) (s_{3}   n_{3} -  b_{3})
  -(s_{3}   n_{2} + a_{3}  n_{1}) (s_{2}    n_{3}  - a_{2}n_{1})] -
\nonumber
\\
- (s_{1} n_{2} - a_{1} n_{3})
[  (s_{2}    n_{1}  + a_{2}n_{3}) (s_{3}   n_{3} -  b_{3}) -
   (s_{3}   n_{1}  - a_{3}  n_{2}) (s_{2}    n_{3}  - a_{2}n_{1}) ] +
\nonumber
\\
   + (s_{1} n_{3} + a_{1} n_{2})
   [ (s_{2}    n_{1}  + a_{2}n_{3}) (s_{3}   n_{2} + a_{3}  n_{1}) -
     (s_{3}   n_{1}  - a_{3}  n_{2} ) (s_{2}    n_{2}  - b_{2}) ] =
\nonumber
\nonumber
\end{eqnarray}
\begin{eqnarray}
=
(s_{1} n_{1} -  b_{1})
[  s_{2} s_{3}   n_{2} n_{3} - s_{2}n_{2} b_{3}   - s_{3}n_{3} b_{2} + b_{2}b_{3} -
s_{2} s_{3}   n_{2} n_{3}  +  s_{3} n_{1} n_{2} a_{2} - n_{1} n_{3} a_{3} s_{2} + a_{2} a_{3} n_{1}^{2} ] -
\nonumber
\\
- (s_{1} n_{2} - a_{1} n_{3})
[ s_{2} s_{3} n_{1} n_{3}  - s_{2} n_{1} b_{3}    + a_{2} s_{3} n_{3}^{2}  -a_{2} n_{3} b_{3}  -
s_{2} s_{3} n_{1} n_{3}  + s_{3} a_{2} n_{1}^{2} + a_{3} s_{2}  n_{2} n_{3} - a_{3} a_{2} n_{1} n_{2}   ] +
\nonumber
\\
   + (s_{1} n_{3} + a_{1} n_{2})
   [ s_{2} s_{3} n_{1} n_{2} +s_{2} a_{3} n_{1}^{2}   + a_{2} s_{3} n_{2} n_{3} +a_{2}a_{3} n_{1} n_{3} -
   s_{2} s_{3} n_{1} n_{2}  + s_{3}b_{2} n_{1} + a_{3} s_{2} n_{2}^{2}  -a_{3} b_{2} n_{2}    ]
\nonumber
\end{eqnarray}
\begin{eqnarray}
=
   - s_{1} n_{1}  s_{2}n_{2} b_{3}   -  s_{1} n_{1} s_{3}n_{3} b_{2} +
s_{1} n_{1} b_{2}b_{3}  +   s_{1} n_{1} s_{3} n_{1} n_{2} a_{2} -  s_{1} n_{1} n_{1} n_{3} a_{3} s_{2}
+ s_{1} n_{1} a_{2} a_{3} n_{1}^{2} -
            \nonumber
            \\
  + b_{1} s_{2}n_{2} b_{3}   + b_{1} s_{3}n_{3} b_{2} - b_{1} b_{2}b_{3}
            -b_{1}   s_{3} n_{1} n_{2} a_{2} + b_{1} n_{1} n_{3} a_{3} s_{2} - b_{1} a_{2} a_{3} n_{1}^{2}
 +
\nonumber
\\[2mm]
 + s_{1} n_{2} s_{2} n_{1} b_{3}    - s_{1} n_{2}  a_{2} s_{3} n_{3}^{2}  + s_{1} n_{2}a_{2} n_{3} b_{3} +
s_{1} n_{2}  s_{3} a_{2} n_{1}^{2} - s_{1} n_{2} a_{3} s_{2}  n_{2} n_{3} + s_{1} n_{2} a_{3} a_{2} n_{1} n_{2}-
\nonumber
\\
 -  a_{1} n_{3} s_{2} n_{1} b_{3}    + a_{1} n_{3} a_{2} s_{3} n_{3}^{2}  - a_{1} n_{3} a_{2} n_{3} b_{3}
+  a_{1} n_{3} s_{3} a_{2} n_{1}^{2} +  a_{1} n_{3} a_{3} s_{2}  n_{2} n_{3} -  a_{1} n_{3} a_{3} a_{2} n_{1} n_{2}    +
\nonumber
\\[2mm]
  + s_{1} n_{3} s_{2} a_{3} n_{1}^{2}   +  s_{1} n_{3} a_{2} s_{3} n_{2} n_{3} + s_{1} n_{3} a_{2}a_{3} n_{1} n_{3}
     + s_{1} n_{3} s_{3}b_{2} n_{1} + s_{1} n_{3} a_{3} s_{2} n_{2}^{2}  - s_{1} n_{3} a_{3} b_{2} n_{2} +
     \nonumber
     \\
      + a_{1} n_{2} s_{2} a_{3} n_{1}^{2}   + a_{1} n_{2} a_{2} s_{3} n_{2} n_{3} + a_{1} n_{2} a_{2}a_{3} n_{1} n_{3}
     + a_{1} n_{2} s_{3}b_{2} n_{1} +  a_{1} n_{2} a_{3} s_{2} n_{2}^{2}  - a_{1} n_{2} a_{3} b_{2} n_{2}
 \nonumber
 \end{eqnarray}

\noindent and further
\begin{eqnarray}
0=  -  b_{1} b_{2}b_{3}   n_{1}^{2} n_{3}^{2}  +
 b_{1} b_{2}b_{3}   n_{1}^{2}  +    b_{1} b_{3} a_{2} n_{1}^{3} n_{2} n_{3}
 -  b_{1}   b_{2} a_{3}  n_{1}^{3} n_{2} n_{3}
+ b_{1} a_{2} a_{3}  n_{1} ^{4}  -
            \nonumber
            \\
  + b_{1} b_{2} b_{3}  n_{2}^{2}     + b_{1} b_{2}  b_{3} n_{3}^{2}  - b_{1} b_{2}b_{3}
            - b_{1}   b_{3}  a_{2} n_{1} n_{2}n_{3}   +
               b_{1} b_{2} a_{3}  n_{1} n_{2} n_{3} - b_{1} a_{2} a_{3} n_{1}^{2}
 +
\nonumber
\\[2mm]
     - b_{1} b_{3}  a_{2}   n_{1} n_{2}   n_{3}^{3}  + b_{1} b_{3} a_{2} n_{1} n_{2}  n_{3}  +
b_{1} b_{3} a_{2} n_{2}   n_{3} n_{1}^{3} - b_{1} b_{2} a_{3}  n_{1}  n_{2}    n_{2}^{2} n_{3} +
b_{1}  a_{3} a_{2} n_{1}^{2} n_{2}^{2} -
\nonumber
\\
 -  b_{2} b_{3} a_{1} n_{3}  n_{2}  n_{1}     + a_{1}  b_{3}  a_{2}  n_{3}^{4}  - a_{1} b_{3}  a_{2} n_{3}^{2}
+  b_{3} a_{1} a_{2}  n_{3}^{2}   n_{1}^{2} +   b_{2} a_{1}  a_{3}   n_{2}^{2} n_{3}^{2} -
a_{1} a_{2} a_{3}   n_{1} n_{2}  n_{3}   +
\nonumber
\\[2mm]
  + b_{1} b_{2} a_{3}    n_{2} n_{3}    n_{1}^{3}   +  b_{1} a_{2} b_{3}    n_{3}^{3} n_{1}  n_{2}
   + b_{1} a_{2} a_{3} n_{1}^{2}  n_{3}^{2}
     + b_{1}  b_{2} b_{3}  n_{3}^{2} n_{1}^{2} + b_{1} b_{2} a_{3}  n_{1} n_{3} n_{2}^{3}  -
     b_{1} a_{3} b_{2}  n_{1}   n_{2} n_{3}  +
     \nonumber
     \\
      + a_{1}b_{2}a_{3}   n_{2} ^{2} n_{1}^{2}   + a_{1}a_{2} b_{3}  n_{2}^{2} n_{3}^{2} +
      a_{1}a_{2} a_{3}   n_{1}  n_{2}  n_{3}
     + a_{1}b_{2}b_{3}   n_{1} n_{2}  n_{3}   +  a_{1} a_{3} b_{2}   n_{2}^{4}  -
     a_{1}  a_{3} b_{2} n_{2}^{2} \equiv = 0  \; ;
 \nonumber
 \end{eqnarray}

\noindent
it is easily verified that all terms cancel out each other indeed.

Let let us consider one other example and  start with a complex  scalar plane  wave:
\begin{eqnarray}
\Phi (x) = e^{i(k_{0}x_{0} - k_{j} x_{j})} \; , \qquad  L_{c} = k_{c} \cos \varphi\;, \;\;
K_{c} = k_{c} \sin \varphi\;;
\nonumber
\end{eqnarray}

\noindent
and eqs. (\ref{A4}) take the form
\begin{eqnarray}
-(a_{0} k_{0} + b_{j}k_{j} )\; k_{c} \cos \varphi  +
(b_{0} k_{0} - a_{j}k_{j} )\; k_{c} \sin \varphi = 0 \; ;
\nonumber
\\
+(b_{0} k_{0} - a_{j}k _{j} )\; k_{c} \cos \varphi +
(a_{0} k_{0} + b_{j} k_{j} )\; k_{c}  \sin \varphi= 0 \; .
\nonumber
\end{eqnarray}

\noindent
from whence it follow
\begin{eqnarray}
 b_{0} k_{0} + ( a_{1} k_{1} + a_{2} k_{2} + a_{3}
 k_{3} )  = 0 \; , \qquad
 a_{0} k_{0}      - ( b_{1}  k_{1}  +
b_{2} k_{2} + b_{3}  k_{3}   = 0 \; .
\nonumber
\end{eqnarray}

\noindent or
\begin{eqnarray}
 b_{0}  = -( a_{1} n_{1} + a_{2} n_{2} + a_{3}  n_{3} ) \; , \qquad
 a_{0}= + ( b_{1}  n_{1}  + b_{2} n_{2} + b_{3}  n_{3}) \; .
\label{5.20}
\end{eqnarray}

\noindent Correspondingly, eq. (\ref{5.8}) gives
\begin{eqnarray}
\Psi (x) =
[  + ( b_{1}  n_{1}  + b_{2} n_{2} + b_{3}  n_{3}) -
i  ( a_{1} n_{1} + a_{2} n_{2} + a_{3}  n_{3} ) ]\;   \Psi^{0}  +
\nonumber
\\
+  (a_{1} + i b_{1}) \Psi^{1} +
(a_{2} + i b_{2}) \Psi^{2} +  (a_{3} + i b_{3}) \Psi^{3}  =
\left |
\begin{array}{c} 0 \\ {\bf E} + i c{\bf B}
\end{array} \right |
\nonumber
\end{eqnarray}

\noindent
where (see (\ref{5.1}) )
\begin{eqnarray}
\{ \Psi^{0} ,  \Psi^{1}, \Psi^{2} ,\Psi^{3} \} =
\left | \begin{array}{rrrr}
i  &  \quad  -n_{1} &  \quad  -n_{2}  &  \quad  -n_{3}   \\
 n_{1}  &  \quad i &  \quad n_{3} &  \quad  -n_{2}  \\
 n_{2}  &  \quad  -n_{3} &  \quad i  &  \quad n_{1}   \\
 n_{3}  &  \quad n_{2} &  \quad -n_{1} &  \quad i
\end{array}  \right |  \;  i k_{0} A \; ( \cos \varphi + i \sin \varphi  ) ,  \; .
\nonumber
\end{eqnarray}

Further we get
\begin{eqnarray}
\Psi = [  -i ( {\bf a} +i {\bf b}) \; {\bf n} \; \Psi^{0} + (a_{1} + i b_{1}) \Psi^{1} +
(a_{2} + i b_{2}) \Psi^{2} +  (a_{3} + i b_{3}) \Psi^{3}  ]=
\nonumber
\\
=
\left | \begin{array}{r}
   ( {\bf a} +i {\bf b})  {\bf n}      - (a_{1} + i b_{1}) n_{1} -  (a_{2} + i b_{2}) n_{2} - (a_{3} + i b_{3}) n_{3}   \\
 - i ( {\bf a} +i {\bf b}) {\bf n} \; n_{1} + (a_{1} + i b_{1}) i     +  (a_{2} + i b_{2}) n_{3} - (a_{3} + i b_{3}) n_{2}  \\
 - i ( {\bf a} +i {\bf b}) {\bf n}\;  n_{2} - (a_{1} + i b_{1}) n_{3} +  (a_{2} + i b_{2}) i     + (a_{3} + i b_{3}) n_{1}   \\
 - i ( {\bf a} +i {\bf b}) {\bf n}\;  n_{3} + (a_{1} + i b_{1}) n_{2} -  (a_{2} + i b_{2}) n_{1} + (a_{3} + i b_{3}) i
\end{array}  \right |  \;  i k_{0} A \; ( \cos \varphi + i \sin \varphi  ) ,  \; .
\nonumber
\end{eqnarray}

\noindent
that is
\begin{eqnarray}
\Psi =
\left | \begin{array}{c}
 0   \\
( {\bf a} +i {\bf b}) {\bf n} n_{1} - (a_{1} + i b_{1})        +  i(a_{2} + i b_{2}) n_{3} - i(a_{3} + i b_{3}) n_{2}  \\
( {\bf a} +i {\bf b}) {\bf n} n_{2} - i(a_{1} + i b_{1}) n_{3} -  (a_{2} + i b_{2})      + i(a_{3} + i b_{3}) n_{1}   \\
( {\bf a} +i {\bf b}) {\bf n} n_{3} + i(a_{1} + i b_{1}) n_{2} -  i(a_{2} + i b_{2}) n_{1} - (a_{3} + i b_{3})
\end{array}  \right |     k_{0} A \; ( \cos \varphi + i \sin \varphi  )   \;
\nonumber
\end{eqnarray}

\noindent which is equivalent to
\begin{eqnarray}
\Psi =
\left | \begin{array}{c}
0  \\
({\bf a} {\bf n} n_{1} - a_{1} - b_{2}n_{3} + b_{3}n_{2}) + i ( {\bf b} {\bf n} n_{1} - b_{1} - n_{2}a_{3}+n_{3}a_{2} \\
({\bf a} {\bf n} n_{2} - a_{2} - b_{3}n_{1} + b_{1}n_{3}) + i ( {\bf b} {\bf n} n_{2} - b_{2} - n_{3}a_{1}+n_{1}a_{2} \\
({\bf a} {\bf n} n_{3} - a_{3} - b_{1}n_{2} + b_{2}n_{1}) + i ( {\bf b} {\bf n} n_{3} - b_{3} - n_{1}a_{2}+n_{2}a_{1}
\end{array} \right | k_{0} A \; ( \cos \varphi + i \sin \varphi  ) \; .
\label{5.21}
\end{eqnarray}

\noindent With the use of notation
\begin{eqnarray}
{\bf L}  = {\bf n} \; ({\bf n}{\bf a}) - {\bf a} - {\bf b } \times {\bf n} \; ,
\qquad
{\bf C} = {\bf n} \; ({\bf n}{\bf b}) - {\bf b} + {\bf a } \times {\bf n} ,
\nonumber
\end{eqnarray}

\noindent
relationship (\ref{5.21}) can be written shorter
\begin{eqnarray}
{\bf E} + i c {\bf B} = ( {\bf L}  + i {\bf C} )\; k_{0} A \; ( \cos \varphi + i \sin \varphi  ) \; ,
\label{5.22}
\end{eqnarray}

\noindent
from whence it follow
\begin{eqnarray}
{\bf E} =  k_{0} A   \; (  \cos \varphi   {\bf L}    -   \sin \varphi {\bf C}  )\; ,
\qquad
c{\bf B} =  k_{0} A  \; (  \sin \varphi   {\bf L}    +   \cos \varphi {\bf C}  ) \; .
\label{5.23}
\end{eqnarray}

\noindent
One can readily prove identities:
\begin{eqnarray}
{\bf L}^{2} = {\bf C}^{2} =
 {\bf a}^{2}  + {\bf b}^{2} - ({\bf n}{\bf a})^{2}
- ({\bf n}{\bf b})^{2} +2 {\bf n} \; ({\bf a} \times {\bf b} ) \; ,
\nonumber
\\
{\bf L } \; {\bf C} = 0 \; , \qquad
{\bf E} {\bf B} = 0 \; , \qquad {\bf E}^{2} = c^{2} {\bf B}^{2} \; ,
\nonumber
\\
{\bf L} {\bf n} = 0 \; ,  \qquad {\bf C} {\bf n} = 0  ,  {\bf E} {\bf n} = 0  \; , \qquad {\bf B} {\bf n} = 0 \; .
\label{5.24}
\end{eqnarray}

General expressions for ${\bf L}, {\bf C}$ may be decomposed into the sum:
\begin{eqnarray}
{\bf L} = {\bf L}_{1} + {\bf  L}_{2} \; , \qquad {\bf C} = {\bf C}_{1} + {\bf  C}_{2} \; ,
\nonumber
\\
I\qquad \underline{{\bf b} = 0:} \qquad
{\bf L}_{1}  = {\bf n} \; ({\bf n}{\bf a}) - {\bf a}  \; , \qquad
{\bf C} _{1} =   {\bf a } \times {\bf n} \; ,
\nonumber
\\[2mm]
{\bf E}_{1}  \times c{\bf B}_{1}  = k_{0}^{2} A^{2} \; ({\bf L_{1}} \times {\bf C}_{1}) =
 k_{0}^{2} A^{2} \; [\; a^{2} - ({\bf n} {\bf a})^{2} \; ] \;  {\bf n}\; ;
\nonumber
\end{eqnarray}
\begin{eqnarray}
II \qquad \underline{{\bf a} = 0:} \qquad
{\bf L}_{2}   =  - {\bf b } \times {\bf n} \; , \qquad
{\bf C}_{2}  = {\bf n} \; ({\bf n}{\bf b}) - {\bf b}  \; ,
\nonumber
\\[2mm]
{\bf E}_{2}  \times c{\bf B}_{2} = k_{0}^{2} A^{2} \; ({\bf L_{2}} \times {\bf C}_{2}) =
k_{0}^{2} A^{2} \;   [\; b^{2} - ({\bf n} {\bf b})^{2} \; ] \;  {\bf n}\; ,
\label{5.25}
\end{eqnarray}

\noindent
In other words, these two  electromagnetic wave are clockwise polarized.

For a particular case when ${\bf b} = {\bf a}$, we get
\begin{eqnarray}
I\qquad
{\bf L}_{1}  = {\bf n} \; ({\bf n}{\bf a}) - {\bf a}  \; , \qquad
{\bf C}_{1} =   {\bf a } \times {\bf n} \; ,
\nonumber
\\
{\bf E}_{1}  = k_{0}A ( \cos \varphi \; {\bf L}_{1} - \sin \varphi \; {\bf C}_{1} )\; ,
\nonumber
\\
c{\bf B}_{1} =  k_{0} A  \; (  \sin \varphi   {\bf L}_{1}    +   \cos \varphi {\bf C}_{1}  ) \; ;
\nonumber
\\
II \qquad
{\bf L} _{2} =  - {\bf a } \times {\bf n} = -{\bf C}_{1}  \; , \qquad
{\bf C} _{2}= {\bf n} \; ({\bf n}{\bf a}) - {\bf a}  = {\bf L}_{1}\; ,
\nonumber
\\
{\bf E}_{2}  = k_{0}A (  - \sin \varphi \; {\bf L}_{1} - \cos \varphi \; {\bf C}_{1} )\; ,
\nonumber
\\
c{\bf B}_{2} =  k_{0} A  \; (  -\sin \varphi   {\bf C}_{1}    +   \cos \varphi {\bf L} _{1} ) \; .
\label{6.26}
\end{eqnarray}

\noindent so constructed waves are linearly independent  and orthogonal:
\begin{eqnarray}
{\bf E} _{1} {\bf E}_{2} = 0 \; , \qquad  {\bf B} _{1} {\bf B}_{2} = 0 \; .
\end{eqnarray}

\noindent
In view of linearity of the Maxwell equations any linear combination of the type is a solution as well:
\begin{eqnarray}
{\bf E} =  c_{1} \; {\bf E}_{1} + c_{2}  \; {\bf E}_{2}  \; , \qquad
{\bf B} =  c_{1} \; {\bf B}_{1} + c_{2}  \; {\bf B}_{2}  \; .
\label{6.27}
\end{eqnarray}

\section{Cylindrical waves}

Let us start with a cylindrical scalar wave  (below for brevity $k_{0}= E$):
\begin{eqnarray}
\Phi =  e^{i E x_{0}}  \;  e^{i k  z }  \; e^{i m \phi } \; R(\rho)  \; ,
\label{6.1}
\\[3mm]
x_{1} = \rho \; \cos \phi \; , \; x_{2} = \rho \; \sin \phi \; , \; x_{3} =  z \; ,
\nonumber
\\
{\partial \over \partial x_{1}} = {\partial \rho \over \partial x_{1}} {\partial \over \partial \rho} +
{\partial \phi  \over \partial x_{2}} {\partial \over \partial \phi} =
\cos \phi \;  {\partial \over \partial \rho} -
{\sin \phi \over \rho}\;  {\partial \over \partial \phi} \; ,
\nonumber
\\
{\partial \over \partial x_{2}} = {\partial \rho \over \partial x_{2} } {\partial \over \partial \rho} +
{\partial \phi  \over \partial x_{2} } {\partial \over \partial \phi} =
\sin \phi   \; {\partial \over \partial \rho} +
{\cos \phi \over  \rho }  \; {\partial \over \partial \phi}  \; .
\label{6.2}
\end{eqnarray}

Corresponding electromagnetic solutions are to be constructed on the base of relations:
\begin{eqnarray}
\{ \Psi^{0} ,  \Psi^{1}, \Psi^{2} ,\Psi^{3} \} =
\lambda _{0} \Psi^{0}  +  \lambda _{1} \Psi^{1} + \lambda _{2} \Psi^{2} + \lambda _{3}\Psi^{3}  =
\left |
\begin{array}{c} 0 \\ {\bf E} + i c{\bf B}
\end{array} \right |.
\label{6.3}
\end{eqnarray}

Let us find $F_{a}$:
\begin{eqnarray}
F_{0}=
iE \; e^{i E x_{0}}  \;  e^{i k  z }  \; e^{i m \phi } \; R(\rho) \; , \qquad
F_{3} = ik \; e^{i E x_{0}}  \;  e^{i k  z }  \; e^{i m \phi } \; R(\rho) \; ,
\nonumber
\\
F_{1}
=
 e^{i E x_{0}}  \;  e^{i k  z }  \; e^{i m \phi } \;
(\cos \phi \;  {d \over d \rho }  - i m \;
{\sin \phi \over \rho}    ) \; R  \; ,
\nonumber
\\
F_{2}
=   e^{i E x_{0}}  \;  e^{i k  z }  \; e^{i m \phi } \;
(\sin \phi \;  {d \over d \rho }  + i m \;
{\cos \phi \over \rho}  ) \;  R \; ,
\label{6.4}
\end{eqnarray}

The main equations to solve are
\begin{eqnarray}
(-\lambda_{0} \partial_{0} + i \lambda_{j} \partial_{j} ) \; F_{c} = 0
\label{B1}
\end{eqnarray}

\noindent
or
\begin{eqnarray}
[\; -i \lambda_{0} E  - \lambda_{3}  k +
i \lambda_{1} ( \cos \phi \;  {\partial \over \partial \rho} -
{\sin \phi \over \rho}\;  {\partial \over \partial \phi}) +
i \lambda_{2} ( \sin \phi   \; {\partial \over \partial \rho} +
{\cos \phi \over  \rho }  \; {\partial \over \partial \phi} ) \; ]\; F_{c} =0 \; ,
\nonumber
\end{eqnarray}

\noindent
or
\begin{eqnarray}
[\; -i \lambda_{0} E  - \lambda_{3}  k +
i ( \lambda_{1} \cos \phi  +  \lambda_{2} \sin \phi ) \;  {\partial \over \partial \rho}
+ i {  - \lambda _{1} \sin \phi +  \lambda_{2} \cos \phi \over  \rho } {\partial \over \partial \phi}
 \; ]\; F_{c} =0 \; .
\label{B1'}
\end{eqnarray}

Let $c=1$:
\begin{eqnarray}
[ -i \lambda_{0} E  - \lambda_{3}  k +
i ( \lambda_{1} \cos \phi  +  \lambda_{2} \sin \phi ) \;  {\partial \over \partial \rho}
+
\nonumber
\\
+  i {  - \lambda _{1} \sin \phi +  \lambda_{2} \cos \phi \over  \rho } {\partial \over \partial \phi}
  ]   e^{i m \phi } \;
(\cos \phi   {d \over d \rho }  - i m
{\sin \phi \over \rho}    )  R  =0  ,
\nonumber
\end{eqnarray}

Let $c=2$:
\begin{eqnarray}
[ -i \lambda_{0} E  - \lambda_{3}  k +
i ( \lambda_{1} \cos \phi  +  \lambda_{2} \sin \phi ) \;  {\partial \over \partial \rho}
+
\nonumber
\\
+ i {  - \lambda _{1} \sin \phi +  \lambda_{2} \cos \phi \over  \rho } {\partial \over \partial \phi}
  ]   e^{i m \phi } \;
(\sin \phi   {d \over d \rho }  + i m
{\cos \phi \over \rho}    )  R  =0  ,
\nonumber
\end{eqnarray}

\noindent
Noting that
\begin{eqnarray}
F_{1} + i F_{2} = e^{i E x_{0}}  \;  e^{i k  z }  \; e^{i (m +1)\phi } ({d \over d \rho } -{m \over \rho }) R \; ,
\nonumber
\\
F_{1} - i F_{2} = e^{i E x_{0}}  \;  e^{i k  z }  \; e^{i (m -1)\phi } ({d \over d \rho } +{m \over \rho }) R \; ,
\label{B2}
\end{eqnarray}

\noindent
from previous  two equations we get
\begin{eqnarray}
[ -i \lambda_{0} E  - \lambda_{3}  k +
i ( \lambda_{1} \cos \phi  +  \lambda_{2} \sin \phi ) \;  {\partial \over \partial \rho}
+  {   \lambda _{1} \sin \phi -  \lambda_{2} \cos \phi \over  \rho } \;  (m +1) \;
  ]    \;
({d \over d \rho }  -
{m \over \rho}    )  R  =0 \;  ,
\nonumber
\end{eqnarray}
\begin{eqnarray}
[ -i \lambda_{0} E  - \lambda_{3}  k +
i ( \lambda_{1} \cos \phi  +  \lambda_{2} \sin \phi ) \;  {\partial \over \partial \rho}
+  {   \lambda _{1} \sin \phi -  \lambda_{2} \cos \phi \over  \rho } \;  (m -1)
  \; ]    \;
({d \over d \rho }  +
{m \over \rho}    )  R  =0  \; .
\label{B3}
\end{eqnarray}

In turn, when $c=0, 3$, we get one the same equation:
\begin{eqnarray}
[\; -i \lambda_{0} E  - \lambda_{3}  k +
i ( \lambda_{1} \cos \phi  +  \lambda_{2} \sin \phi ) \;  {\partial \over \partial \rho}
+  {   \lambda _{1} \sin \phi -  \lambda_{2} \cos \phi \over  \rho } \; m \; ]\; R(\rho)  =0 \; ,
\label{B4}
\end{eqnarray}

We  readily note that there exist very simple way to satisfy  these three equations on parameters $\lambda_{c}$:
\begin{eqnarray}
 -i \lambda_{0} E  - \lambda_{3}  k = 0 \;, \qquad \lambda_{1} = 0\; , \qquad \lambda_{2} = 0 \; .
 \label{B5}
 \end{eqnarray}

Let us  demonstrate that no other solutions exist.
To this end, with notation
\begin{eqnarray}
C =    -i \lambda_{0} E  - \lambda_{3}  k   \; , \qquad
A =   \lambda_{1} \cos \phi  +  \lambda_{2} \sin \phi  \; , \qquad
B =     \lambda _{1} \sin \phi -  \lambda_{2} \cos \phi  \; .
\nonumber
\end{eqnarray}

\noindent let us rewrite  eqs. (\ref{B3})  --  (\ref{B4})   in the form
\begin{eqnarray}
[\; C + i A  {d \over d \rho} + {B \over  \rho } \;   (m+1) \; ]\;  ({dR \over d \rho }
 -{m \over \rho }  \; R  )= 0 \; ,
\nonumber
\\
\; [\; C + i A {d \over d \rho  }+ {B \over  \rho } \;  (m-1) \; ]\;
({dR \over d \rho } +{m \over \rho }\;  R )  = 0 \; ,
\nonumber
\\
\; [ \;  C + i A  {d \over d \rho } + {B \over  \rho }  m  \; ]\; R = 0\; .
\label{B6}
\end{eqnarray}

\noindent
Combining two first equations in (\ref{B6})(summing and subtracting),  we get
\begin{eqnarray}
C {d R \over d \rho } + i A \; {d^{2} R \over d \rho^{2}} +  {mB \over \rho }  {d R \over d \rho } -
 { m B \over \rho^{2} } R = 0 \; ,
\nonumber
\\
- {mC \over \rho} R -  i m A \; {d \over d \rho } {R \over  \rho}  +
 {B \over \rho } {d R \over d \rho} -  m^{2} {B \over \rho^{2}} R = 0 \; ,
\nonumber
\\
   C \; R + i A \; {d  R \over d \rho } +  m  \; {B \over  \rho }  \;  R = 0 \; .
\label{B7}
\end{eqnarray}

After differentiating  the third  equation will look
\begin{eqnarray}
 C \; {d R\over d \rho }  + i A \; {d^{2}  R \over d \rho^{2}  }
  +    {mB \over \rho } {d R \over d \rho}  -    { mB \over  \rho ^{2} }  \;  R = 0
\nonumber
\end{eqnarray}

\noindent which coincides with the first equation in (\ref{B7}). Therefore, the system (\ref{B7}) is equivalent to
\begin{eqnarray}
- {mC \over \rho} R -  i m A \; {d \over d \rho } {R \over  \rho}  +
 {B \over \rho } {d R \over d \rho} -  m^{2} {B \over \rho^{2}} R = 0 \; ,
\nonumber
\\
   C \; R + i A \; {d  R \over d \rho } +  m  \; {B \over  \rho }  \;  R = 0 \; .
\label{B8}
\end{eqnarray}

\noindent
The system  (\ref{6.17}) can be rewritten as follows:
\begin{eqnarray}
-m \;{C \over \rho}\; R  +  {B \over \rho }\; {d R \over d \rho}
+ i  \;{m A \over  \rho^{2} } \; R  -  {m \over \rho } \;( i A  \; {d  R \over d \rho }
 + m \; {B \over \rho} \;R ) = 0 \; ,
 \nonumber
\\
   C \; R + i A \; {d  R \over d \rho } +  m  \; {B \over  \rho }  \;  R = 0 \; .
\nonumber
\end{eqnarray}

\noindent here the first equation  with the use of the second equation gives
\begin{eqnarray}
 {B \over \rho }\; {d R \over d \rho}
+ i  \;{m A \over  \rho^{2} } \; R
=0 \; .
\nonumber
\end{eqnarray}

Therefore system (\ref{B8}) is equivalent to
\begin{eqnarray}
 B \; {d R \over d \rho}
+ i  \;{m A \over  \rho } \; R
=0 \; ,
 \nonumber
\\
   i A \; {d  R \over d \rho } +  m  \; {B \over  \rho }  \;  R  + C \; R  = 0 \;
\nonumber
\end{eqnarray}

\noindent
which in turn is equivalent to
\begin{eqnarray}
(B +iA) ({d \over d \rho } +{m \over  \rho}) R + C\; R =0 \; ,
\nonumber
\\
(B-iA) ({d \over d \rho } - {m \over  \rho}) R - C\; R =0 \; ,
\label{B9}
\end{eqnarray}

Noting identity
$(A+iB)(A-iB) = \lambda_{1}^{2} + \lambda_{2}^{2} \;$,
one reduces eqs. (\ref{B9}) to the form
\begin{eqnarray}
( \lambda_{1}^{2} + \lambda_{2}^{2} ) ({d \over d \rho } +{m \over  \rho}) R +   (B -iA ) C  \; R =0 \; ,
\nonumber
\\
 ( \lambda_{1}^{2} + \lambda_{2}^{2} ) ({d \over d \rho } -{m \over  \rho}) R +  (B+iA)C  \; R =0 \;.
\label{B10}
\end{eqnarray}

\noindent
Remembering that
\begin{eqnarray}
C =    -i \lambda_{0} E  - \lambda_{3}  k   \; , \qquad
A =   \lambda_{1} \cos \phi  +  \lambda_{2} \sin \phi  \; , \qquad
B =     \lambda _{1} \sin \phi -  \lambda_{2} \cos \phi  \; .
\nonumber
\end{eqnarray}

\noindent
we immediately conclude that  eqs.  (\ref{B10})  can be satisfied only by the following way:
\begin{eqnarray}
( \lambda_{1}^{2} + \lambda_{2}^{2} ) \; , \qquad C = 0 \; .
\label{B11}
\end{eqnarray}

In other words,
\noindent
This means that relations (\ref{B5})  provide us with the only possible solution
 in terms of two
($1 \times 4$) columns
\begin{eqnarray}
\Psi = \lambda_{0} \Psi^{0} + \lambda_{3} \Psi^{3}\;, \qquad
\qquad    \underline{- \lambda_{0}\;  E  +  i \; \lambda _{3} k =0   } \; ,
\nonumber
\\
 \Psi^{0}  =
\left | \begin{array}{r}
i F_{0}(x)     \\
 -F_{1}(x)    \\
 -F_{2}(x)     \\
 -F_{3}(x)
\end{array}  \right | \; , \quad
\Psi^{3}  =
\left | \begin{array}{rrrr}
  F_{3}(x)   \\
  F_{2}(x)  \\
-F_{1}(x)   \\
 iF_{0}(x)
\end{array}  \right | \; ,
\nonumber
\\
F_{0}=
iE \; \Phi \; , \qquad F_{1}
=
 (\cos \phi \;  {d \over d \rho }  - i m \;
{\sin \phi \over \rho}    ) \; \Phi  \; ,
\nonumber
\\
F_{2}
=
(\sin \phi \;  {d \over d \rho }  + i m \;
{\cos \phi \over \rho}  ) \;  \Phi  \; ,
\; , \qquad
F_{3} = ik \; \Phi \; ,
\label{6.21}
\end{eqnarray}

\noindent
Solution explicitly reads by components:
$$
(\Psi)_{0} =
(  - \lambda_{0} \;  E + i\; \lambda_{3}  k  ) \;\Phi = 0 \; ,
$$
$$
E_{3} + i B_{3} = (\Psi)_{3} =
(  - \lambda_{0} \; ik -  \lambda_{3}  E  ) \; \Phi = - \lambda_{3}  \; {E^{2} - k^{2} \over E}  \; \Phi  \; ,
$$
 $$
E_{1} + i B_{1} = (\Psi)_{1} =
\left [  -\lambda_{0} \; (\cos \phi \;  {d \over d \rho }  - i m \;
{\sin \phi \over \rho}    ) \;   + \lambda_{3} \;( \sin \phi \;  {d \over d \rho }  + i m \;
{\cos \phi \over \rho})\; \right  ] \;  \Phi ; ,
$$
$$
E_{2} + i B_{2} = (\Psi)_{2} =   \left [\;
 ( -\lambda_{0} \; (\sin \phi \;  {d \over d \rho }  + i m \;
{\cos \phi \over \rho})  - \lambda_{3} \; ( \cos \phi \;  {d \over d \rho }  - i m \;
{\sin \phi \over \rho}    ) \;  \right ] \; \Phi \; .
$$

\noindent
After simple rewriting we get
$$
E_{3} + i B_{3} = (\Psi)_{3} =  - \lambda_{3}  \; {E^{2} - k^{2} \over E}  \; \Phi  \; ,
$$
$$
E_{1} + i B_{1} = { \lambda _{3} \over E} \;
\left [  - ik  \; (\cos \phi \;  {d \over d \rho }  - i m \;
{\sin \phi \over \rho}    ) \;   + E \;( \sin \phi \;  {d \over d \rho }  + i m \;
{\cos \phi \over \rho})\; \right  ] \;  \Phi \; ,
$$
$$
E_{2} + i B_{2} =   { \lambda _{3} \over E} \;
 \left [\;
 ( - ik  \; (\sin \phi \;  {d \over d \rho }  + i m \;
{\cos \phi \over \rho})  - E  \; ( \cos \phi \;  {d \over d \rho }  - i m \;
{\sin \phi \over \rho}    ) \;  \right ] \; \Phi \; .
$$

For special cases we get most simplicity:
$$
\underline{k = + E \;}  , \qquad
 \qquad
E_{3} + i B_{3} =   0  \; ,
$$
$$
E_{1} + i B_{1} =   \lambda _{3} \;
\left [  - i   \; (\cos \phi \;  {d \over d \rho }  - i m \;
{\sin \phi \over \rho}    ) \;   + ( \sin \phi \;  {d \over d \rho }  + i m \;
{\cos \phi \over \rho})\; \right  ] \;  \Phi \; = -i\;  \lambda_{3} \; \;  e^{i\phi} ( {d \over d \rho } -{m \over  \rho }) \Phi \; ,
$$
$$
E_{2} + i B_{2} =   \lambda _{3}  \;
 \left [\;
 ( - i  \; (\sin \phi \;  {d \over d \rho }  + i m \;
{\cos \phi \over \rho})  -  ( \cos \phi \;  {d \over d \rho }  - i m \;
{\sin \phi \over \rho}    ) \;  \right ] \; \Phi=
- \lambda_{3} \; \;  e^{i\phi} ( {d \over d \rho } -{m \over  \rho }) \Phi \; ,
 \; ;
$$

$$
\underline{k = - E \;}  , \qquad
 \qquad
E_{3} + i B_{3} =  0  \; ,
$$
$$
E_{1} + i B_{1} =  \lambda _{3}  \;
\left [  + i  \; (\cos \phi \;  {d \over d \rho }  - i m \;
{\sin \phi \over \rho}    ) \;   + ( \sin \phi \;  {d \over d \rho }  + i m \;
{\cos \phi \over \rho})\; \right  ] \;  \Phi  =
i\;  \lambda_{3} \; \;  e^{-i\phi} ( {d \over d \rho } +{m \over  \rho })\;\Phi \;  ,
$$
$$
E_{2} + i B_{2} =     \lambda _{3}  \;
 \left [\;
 ( + i   \; (\sin \phi \;  {d \over d \rho }  + i m \;
{\cos \phi \over \rho})  -   ( \cos \phi \;  {d \over d \rho }  - i m \;
{\sin \phi \over \rho}    ) \;  \right ] \; \Phi =
- \lambda_{3} \; \;  e^{-i\phi} ( {d \over d \rho } +{m \over  \rho })\; \Phi \; .
$$

\vspace{5mm}

It seems reasonable to expect further developments in this matrix based approach to Maxwell theory,
 as a possible base to
explore general method to separate the variables for Maxwell equations in different coordinates. Also it would be desirable to extent this approach
to Maxwell theory in curved space -- time models.

\vspace{5mm}

Authors are grateful to all participants of scientific seminar of
Laboratory of Theoretical Physics of Institute of Physic on NASB
for discussion and advice.

\end{document}